\documentclass[10pt,twocolumn,conference]{IEEEtran}
\usepackage{cite}
\ifCLASSINFOpdf
\usepackage[pdftex]{graphicx}
\else
\usepackage[dvips]{graphicx}
\fi
\usepackage{amsmath}
\usepackage{subfig}
\usepackage{enumitem}
\usepackage{dblfloatfix}

\usepackage{wrapfig}
\usepackage{mdframed}

\usepackage{url}
\usepackage{hyperref}
\usepackage[displaymath,mathlines,columnwise,switch]{lineno}
\usepackage{bbm}
\usepackage{stmaryrd}
\usepackage{oktheorem}
\usepackage{mathtools}
\usepackage{mathpartir}
\usepackage{okcategory}

\modulolinenumbers[2]
\usepackage{suffix}
\usepackage{amsmath}
\usepackage{okgeneralmath}
\usepackage{oksemantics}

\newcommand\hide[1]{}

\usepackage{colortbl}
\usepackage{graphics}
\usepackage[table]{xcolor}
\definecolor{shade}{RGB}{223,223,223}
\usepackage{tcolorbox}
\newtcbox{\shadebox}{on line,arc=1pt, outer arc=2pt,%
  colback=shade,colframe=shade,boxsep=0pt,%
  left=1pt,right=1pt,top=2pt,bottom=2pt,%
  boxrule=0pt,bottomrule=1pt,toprule=1pt}
\newcommand{\shade}[1]{%
        \shadebox{\ensuremath{#1}}%
}

\newtcbox{\myfigbox}{%
  on line, arc=1pt, outer arc=1pt,colframe=black,
  colback=white,
  boxsep=0pt,
  left=1pt,right=1pt,top=1pt,bottom=1pt,%
  boxrule=.5pt,bottomrule=.5pt,toprule=.5pt}

\newtcbox{\myfigboxshade}{%
  on line, arc=1pt, outer arc=1pt,colframe=black,
  colback=shade,boxsep=0pt,
  left=1pt,right=1pt,top=1pt,bottom=1pt,%
  boxrule=.5pt,bottomrule=.5pt,toprule=.5pt}

\newcommand\shaderow{\rowcolor{shade}}

\newcommand\seclabel[1]{\label{sec:#1}}
\newcommand\secref[1]{Sec.~\ref{sec:#1}}
\newcommand\subseclabel[1]{\label{subsec:#1}}
\newcommand\subsecref[1]{Subsec.~\ref{subsec:#1}}
\newcommand\figlabel[1]{\label{fig:#1}}
\newcommand\figpref{Fig.}

\newcommand\figref[1]{\figpref~\ref{fig:#1}}
\WithSuffix\newcommand\figref*[1]{\ref{fig:#1}}
\newcommand\subfiglabel[1]{\label{subfig:#1}}
\newcommand\subfigref[1]{\figpref~\ref{subfig:#1}}

\newcommand\keyword[1]{\mathbf{#1}}
\newcommand\keywordop[1]{{\mathrel{\keyword{#1}}{}}}
\WithSuffix\newcommand\keywordop*[2][]{{\mathrel{\keyword{#2}_{#1}{}}}}
\newcommand\keywordrel[1]{\mathrel{\keyword{#1}}}
\newcommand\ConcreteCll[1]{\mathtt{#1}}
\newcommand\linkedlist{\ConcreteCll{linked\_list}}
\newcommand\cell{\ConcreteCll{list\_cell}}
\newcommand\data{\ConcreteCll{data}}
\newcommand\boolty{\keyword{bool}}
\newcommand\sref[1]{{\keyword{ref}_{#1}}}

\newcommand\runST{\keyword{runST}}
\newcommand\sunit{\keyword{()}}
\newcommand\strue {\keyword{true}}
\newcommand\sfalse{\keyword{false}}
\newcommand\isnew[1][]{\keywordop{new}}
\newcommand\islet[2]{\keywordop{let}#1=#2\keywordrel{in}}
\newcommand\x{x}
\newcommand\y{y}
\newcommand\cx{\mathtt{x}}

\renewcommand\infer{\vdash}

\newcommand\gdefinedby{\mathrel{::=}}
\newcommand\gor       {\mathrel{\lvert}}

\newcommand\syncat[1]{\mspace{-25mu}\synname{#1}}
\newcommand\synname[1]{\qquad\text{#1}}
\newenvironment{newsyntax}[1][]{%
\(
  \rowcolors{100}{white}{white}%
  \begin{array}[t]{#1*3{l@{}}@{\,}l}
}{
\end{array}
\)%
}

\newcommand\cellnames{\mathbf{S}}
\newcommand\cellsig{\boldsymbol{\Sigma}}
\newcommand\Cll{c}
\newcommand\grounds{\mathbf{G}}
\newcommand\gty{\gamma}
\newcommand\typeof{{\mathrel{\mathit{ctype}}{}}}

\newcommand\lamref[1][\cellsig]{\lambda^{#1}_{\mathrm{ref}}}

\newcommand\locs{\mathbbm{L}}

\newcommand\ty\tau

\WithSuffix\newcommand\s+{+}
\WithSuffix\newcommand\s*{*}
\WithSuffix\newcommand\s1{\keyword{1}}
\WithSuffix\newcommand\s0{\keyword{0}}
\newcommand\sto{\to}
\newcommand\sv{v}
\newcommand\st{t}
\WithSuffix\newcommand\st'{s}
\newcommand\sinj[2][\ty_1 \s+\ty_2]{\mathord{\keyword{inj}}^{#1}_{#2}}
\newcommand\spair[2]{(#1,#2)}
\newcommand\sseq[1]{(#1)}
\newcommand\sstar{()}
\newcommand\sfun[3]{\lambda #1:#2. #3}
\newcommand\szmatch[2]{%
        \mathop{\keyword{match}} #1
        \mathbin{\keyword{with}} \{\}^{#2}%
}
\newcommand\sbmatch[6][]{%
        \mathop{\keyword{match}} #2
        \mathbin{\keyword{with}}#1 \{%
        \sinj[]1 #3 \mapsto #4
        \mathbin{\gor}\,
        \sinj[]2 #5 \mapsto #6
        \}
}
\newcommand\spmatch[3]{\mathop{\keyword{match}} #1\mathbin{\keyword{with}} \spair{#2}{#3}\mapsto }
\newcommand\sset[2]{#1 \mathrel{:=} #2}
\newcommand\sget[1]{\mathop{!}\nolimits#1}
\newcommand\snew[3][]{
        \mathrel{\keyword{letref}_{#1}}#2
        \mathrel{\keyword{in}} #3
}
\WithSuffix\newcommand\snew*[3][]{
        \mathrlap{\mathrel{\keyword{letref}_{#1}}}#2
        \mathrel{\keyword{in}} #3
}

\newcommand\slet[4]{
        \mathrel{\keyword{let}}(#1:#2) = #3
        \mathrel{\keyword{in}}  #4
}
\WithSuffix\newcommand\slet*[3]{
        \mathrel{\keyword{let}}#1 = #2
        \mathrel{\keyword{in}}  #3
}
\newcommand\newsep{,}
\newcommand\newbind[4][]{\sset{(#2 : \sref{#3}#1)}{#4}}
\newcommand\ssnew[1]{\keywordop*[#1]{new}}

\newcommand\w{w}
\newcommand\pfinto{\rightharpoonup_{\mathrm{fin}}}

\newcommand\judge[3][]{\Tys#1\infer_{\w} #2 : #3}
\WithSuffix\newcommand\judge*[3][]{\Tys#1\infer_{\w_{\infer}} #2 : #3}
\newcommand\subst[2]{#1[#2]}
\WithSuffix\newcommand\subst*[1]{#1\coseq}
\newcommand\replace[2]{#1\mapsto#2}
\newcommand\e{\theta}

\newcommand\Tys\Gamma

\newcommand\heap{\boldsymbol{\eta}}
\newcommand\heaps{\mathbf{H}}
\newcommand\bigstep{\Downarrow}
\newcommand\fresh[2]{\mathop{\#_{#1}}\seq{#2}}

\newcommand\ctxeq{\simeq_{\mathrm{ctx}}}
\newcommand\Plugged[5]{\mathcal{C}[#1\infer_{#2}#3,#4:#5]}

\newcommand\+{\oplus}

\newcommand\scat{\mathbbm{C}} 
\newcommand\worlds{\mathbbm{W}}
\newcommand\init{\mathbbm{E}}
\newcommand\p{u}
\newcommand\comma{\downarrow}
\newcommand\h\rho
\newcommand\ih\epsilon

\newcommand\coend[1]{\int^{#1}}
\WithSuffix\newcommand\coend*[1]{\coend{\mathrlap{#1}}}
\newcommand\cocoend[1]{\int_{#1}}
\WithSuffix\newcommand\cocoend*[1]{\cocoend{\mathrlap{#1}}}

\newcommand\commatise{\overline}
\newcommand\moncat[1]{\underline{#1}}
\renewcommand\*\otimes
\renewcommand\.\odot
\newcommand\vact\boxdot
\newcommand\I{I}

\newcommand\monact{\moncat}
\newcommand\compact{\alpha}
\newcommand\trivact{\lambda}
\newcommand\acthom{\multimap}

\newcommand\vals{\mathbf{W}}
\newcommand\pconfs{\mathbf{E}}

\DeclareSymbolFont{symbolsC}{U}{txsyc}{m}{n}
\SetSymbolFont{symbolsC}{bold}{U}{txsyc}{bx}{n}
\DeclareFontSubstitution{U}{txsyc}{m}{n}
\DeclareMathSymbol{\multimapdot}{\mathrel}{symbolsC}{20}

\DeclareMathAlphabet{\mathbbold}{U}{bbold}{m}{n}

\newcommand\powers{\multimapdot}
\newcommand\structure{\underline}
\newcommand\bind{\mathrel{\scalebox{.5}[1]{$>\!\!>\!=$}}}
\newcommand\strength{\mathrm{str}}
\newcommand\dstrength{\mathrm{dstr}}
\newcommand\eval{\mathrm{eval}}
\newcommand\mreturn{\mathrm{return}}
\newcommand\curry{\mathop{\mathrm{curry}}\nolimits}
\newcommand\uncurry{\mathop{\mathrm{uncurry}}\nolimits}
\newcommand\swap{\mathop{\mathrm{swap}}\nolimits}
\newcommand\ord{\underline}
\newcommand\inject{\rightarrowtail}
\newcommand\lnum{\mathop\#}
\WithSuffix\newcommand\set*[1]{\{#1\}}
\WithSuffix\newcommand\suchthat*{\vert}
\WithSuffix\newcommand\seq*[1]{\mathopen<#1\mathclose>}
\newcommand\numof[1]{\cardinality{#1}}
\newcommand\iinj{\iota^{\+}}
\WithSuffix\newcommand\o-\ominus
\newcommand\comp[1]{#1^{\complement}}
\WithSuffix\newcommand\o+[1][\w]{\+_{#1}}
\newcommand\mis{^\star}
\WithSuffix\newcommand\mis'{^*}

\newcommand\cdiagram[2][c]{\[\cdiagram*[#1]{#2}\]}
\WithSuffix\newcommand\cdiagram*[2][c]{\begin{aligned}[#1]\includegraphics{#2.mps}\end{aligned}}

\newcommand\sem{\scottbrackets}
\renewcommand\terminal{\mathbbm{1}}
\renewcommand\initial{\mathbbold{O}}

\newcommand\stores{\mathbbm{H}}
\newcommand\prostores{\underline\stores}
\newcommand\storesmon{\stores^\+}
\newcommand\storesuni{\stores^{\emptyset}}
\newcommand\store{\eta}

\newcommand\vto[3]{%
  \begin{smallmatrix}%
        \hphantom{#2}#1     \\%
        #2\mathord\downarrow\\%
        \hphantom{#2}#3       %
  \end{smallmatrix}%
}

\newcommand\mget{\mathrm{get}}
\newcommand\mset{\mathrm{set}}
\newcommand\mnew{\mathrm{new}}
\newcommand\mshift{\partial}
\newcommand\minit{\mathrm{init}}
\newcommand\mhide{\mathrm{hide}}

\newcommand\me{e}

\newcommand\vsem[1]{\sem{#1}^{\mathrm v}}
\newcommand\cvsem[1]{\sem{#1}^{\mathrm v}_{\star}}
\newcommand\csem[1]{\sem{#1}_{\star}}

\newcommand\smv{\mathtt{v}}

\newcommand\cbveq[7][\ \equiv]{
  \begin{aligned}[t]
  \mathrlap{#3\infer_{#4}}\qquad&\\
  &\begin{array}{*4{@{}l}}
    #5
  \end{array}
  #1
  \begin{array}{*4{@{}l}}
    #6
  \quad\mathrlap{{}: #7}
  \end{array}
  \end{aligned}
}
\WithSuffix\newcommand\cbveq*[7][\ \equiv]{
  \begin{aligned}[t]
  \mathrlap{#2.~#3\infer_{#4}}\qquad&\\
  &\begin{array}{*4{@{}l}}
    #5
  \end{array}
  #1
  \begin{array}{*4{@{}l}}
    #6
  \end{array}\\
    \mathrlap{{}: #7}\quad&
  \end{aligned}
}
\newcommand{\HidingAlgs}{\mathbf{HidAlg}}

\newcommand\patternA[2]{
  \draw[fill=blue, semitransparent] #1 rectangle #2;
  \draw[pattern=grid, pattern color=cyan, semitransparent] #1 rectangle #2;
}
\WithSuffix\newcommand\patternA*[2]{
  \draw[pattern=grid, pattern color=blue, nearly transparent] #1 rectangle #2;
}
\newcommand\patternB[2]{
  \draw[pattern=crosshatch dots, semitransparent] #1 rectangle #2;
  \draw[fill=red, semitransparent] #1 rectangle #2;
}
\WithSuffix\newcommand\patternB*[2]{
  \draw[pattern=crosshatch dots, semitransparent] #1 rectangle #2;
}

\newcommand\DashedCellUpwards[4][]{
  \node[inner sep=1pt] #2 at #4   {};
  \node[inner sep=1pt] #3 at ($#4 + (1, 1)$) {};
  \node #1 at ($#2!0.5!#3$) {};
  \draw[dashed]
          let \p1 = #2,
              \p2 = #3
          in (\x1,\y1) -- (\x1, \y2) -- (\x2, \y2) -- (\x2, \y1) ;

}
\newcommand\DashedCellDownwards[4][]{
    \node[inner sep=1pt] #2 at #4   {};
    \node[inner sep=1pt] #3 at ($#4 + (1, 1)$) {};
    \node #1 at ($#2!0.5!#3$) {};
    \draw[dashed]
          let \p1 = #2,
              \p2 = #3
          in (\x1, \y2) -- (\x1,\y1) -- (\x2, \y1) -- (\x2, \y2);
}

\newcommand\DashedCellMiddle[4][]{
    \node[inner sep=1pt] #2 at #4   {};
    \node[inner sep=1pt] #3 at ($#4 + (1, 1)$) {};
    \node #1 at ($#2!0.5!#3$) {};
    \draw[dashed]
          let \p1 = #2,
              \p2 = #3
          in (\x1, \y2) -- (\x1,\y1);
    \draw[dashed]
          let \p1 = #2,
              \p2 = #3
          in (\x2, \y1) -- (\x2, \y2);
}

\newcommand\HeapCell[4][]{
    \node[inner sep=1pt] #2 at #4   {};
    \node[inner sep=1pt] #3 at ($#4 + (1,1)$) {};
    \node at ($#2!0.5!#3$) {};
    \draw #2 rectangle #3;
    #1;
}

\newcommand\DashedHeapCell[4][]{
    \node[inner sep=1pt] #2 at #4   {};
    \node[inner sep=1pt] #3 at ($#4 + (1,1)$) {};
    \node at ($#2!0.5!#3$) {};
    \draw[dashed] #2 rectangle #3;
    #1;
}

\usepackage{tikz}
\usetikzlibrary{calc,shapes.multipart,chains,arrows, patterns, cd,matrix}
\usetikzlibrary{arrows.meta}

\begin{document}
\title{A monad for full ground reference cells}

\author{%
  \IEEEauthorblockN{Ohad Kammar%
    \IEEEauthorrefmark{1}%
    \IEEEauthorrefmark{2}%
    \IEEEauthorrefmark{4},
    Paul B.~Levy\IEEEauthorrefmark{3},
    Sean K.~Moss\IEEEauthorrefmark{2}%
        \IEEEauthorrefmark{5}, and
    Sam Staton\IEEEauthorrefmark{1}
  }
  \IEEEauthorblockA{%
    \IEEEauthorrefmark{1}%
    University of Oxford
    Department of Computer Science}
  \IEEEauthorblockA{\IEEEauthorrefmark{2}%
    University of Cambridge %
    \IEEEauthorrefmark{4}Computer Laboratory and
   \IEEEauthorrefmark{5}%
    Department of Pure Mathematics and Mathematical Statistics}
  \IEEEauthorblockA{\IEEEauthorrefmark{3}%
    University of Birmingham
  School of Computer Science}%
}

\IEEEoverridecommandlockouts
\IEEEpubid{\makebox[\columnwidth]{Authors' copy} \hspace{\columnsep}\makebox[\columnwidth]{ }}

\maketitle

\begin{abstract}
We present a denotational account of dynamic allocation of potentially
cyclic memory cells using a monad on a functor category.  We identify
the collection of heaps as an object in a different functor category
equipped with a monad for adding hiding/encapsulation capabilities to
the heaps.  We derive a monad for full ground references supporting
effect masking by applying a state monad transformer to the
encapsulation monad. To evaluate the monad, we present a denotational
semantics for a call-by-value calculus with full ground references, and
validate associated code transformations.




\end{abstract}
\IEEEpeerreviewmaketitle

\section{Introduction}
Linked lists are a common example of a dynamically
allocated, mutable, and potentially cyclic data type involving three
sorts of memory cell.

\begin{itemize}[align=left, itemindent=0pt, listparindent=1em, leftmargin=.5em]
\item $\linkedlist$ cells store values of type $\s1 \s+ \sref
  \cell$, i.e. a variant that is either a \textsc{nil} value or a
  pointer to a $\cell$;
\item $\cell$ cells store values of type ${\sref \data \!\!\s*\! \sref
  \linkedlist}$, i.e.~a pair of pointers to a data payload and another
  list; and
\item $\data$ cells store values of type $\boolty$, i.e., a single bit.
\end{itemize}
Such reference cells, that may contain references to other references
and create cycles, but may not store functions and thunks, are called
\emph{full ground
  references}~\cite{murawski-tzevelekos:algorithmic-games-for-full-ground-references}.

Here we develop a denotational semantics for languages with full
ground storage in which types denote sets-with-structure and program
terms denote structure-preserving functions. When dynamic allocation
is involved, such a semantics typically uses functor categories and is
called possible-world semantics, e.g.~\cite{oles:thesis,
  ohearn-tennent:semantics-of-local-variables, stark:thesis,
  plotkin-power:notions-of-computation-determine-monads}. To motivate
the functorial structure of types, consider the type of $\linkedlist$
and note that unless a memory cell has been previously allocated, the
empty list is the only possible value of this type. As we allocate new
cells, our programs can access more linked lists, and the type of
linked lists is functorial in \emph{worlds}: collections of previously
allocated memory locations.

Our work builds on
Moggi's~\cite{moggi:computational-lambda-calculus-and-monads} theory
of computational effects and
monads. Moggi~\cite{moggi:an-abstract-view-of-programming-languages}
gave monads for dynamic allocation of names and dynamically allocated
\emph{ground} storage, where the range of storable values does not
change between worlds, such as $\sref \data$ cells, which only store a
bit, regardless of what memory locations have been previously
allocated. Plotkin and
Power~\cite{plotkin-power:notions-of-computation-determine-monads}
give a different monad for ground storage on the same functor category
as Moggi. Their monad is Hilbert-Post complete with respect to the
expected first-order program
equivalences~\cite{staton:completeness-for-algebraic-theories-of-local-state}.
Ghica's masters
thesis~\cite{ghica:semantics-of-dynamic-variables-in-algol-like-languages}
pioneered a functor category semantics for Idealised Algol extended
with pointers, where dangling pointers played a crucial role in the
model. The monad we propose here supports full ground storage for
ML-like references without dangling pointers and validates the
expected first-order program equivalences.

Possible-world semantics for local storage typically involves
\emph{heaps}, which assign a value of the appropriate type to each of
a given set of locations. The situation is straightforward in the
(ordinary) ground case, as heaps collect into a functor
\emph{contra}variant in worlds, with the functorial action given by
projection. However, when heaps may contain cyclic data, this
functorial action is ill-defined. For example, projecting a heap
containing a cyclic list with two cells into a world containing only a
single cell reference results in a dangling pointer.

We give a functorial action on heaps by defining the category of
\emph{initialisations}. Its objects are worlds, and its morphisms are
world morphisms together with \emph{initialisation data}: a specified
value for each of the newly added locations in the codomain. The
collection of heaps is functorial with respect to initialisations,
and moreover, there is a way to transform monads over functors from
initialisations into monads over functors from ordinary worlds,
equipping the transformed monad with operations for mutating and dereferencing
pointers.

We choose a specific monad over functors from initialisation which
supports hiding capabilities. Adding this hiding capability to the
heaps object gives the generalisation of the contravariant action
given by projection in the (ordinary) ground case. The monad for full
ground references is given by transforming the hiding monad in the
aforementioned way. We define an allocation operation on this
resulting monad.

To evaluate the suitability of this monad for modelling reference
cells, we use three yardsticks. First, we prove this monad has the
\emph{effect masking} property: the global elements of the monad
applied to a constant functor factor uniquely through the unit of the
monad. We interpret this result as stating that computations that do
not leak references are semantically equivalent to pure values.
Second, we use the monad to give adequate semantics to a total
call-by-value calculus with full ground references. Finally, we show
that the axioms for (non-full) ground
references~\cite{plotkin-power:notions-of-computation-determine-monads,
  staton:completeness-for-algebraic-theories-of-local-state} hold in
this model.


We compare several existing semantic approaches to the denotational
semantics of reference cells, to place the possible-worlds semantics
in context.

\paragraph*{Relational models}
In these
models~\cite{benton-leperchey:relational-reasoning-in-a-nominal-semantics-for-storage,bohr-birkedal:relational-reasoning-for-recursive-types-and-references,
  benton-kennedy-beringer-hofmann:relational-semantics-for-effect-based-program-transformations-with-dynamic-allocation,
  benton-hofmann-nigam:abstract-effects-and-proof-relevant-logical-relations},
heaps may associate to a location values of any type, and the
semantics is given non-functorially. Semantic equivalence in these
models does not validate some of the basic intended equations, such
as:
\[
\islet \cx{\isnew \strue}\strue
\ \equiv\
\strue
\]
These models define, in addition, a logical relations interpretation,
and being related by this relation implies contextual
equivalence~\cite{hofmann:correctness-of-effect-based-program-transformations}. We
then validate the equations of interest with respect to this
relation. Contrasted with the semantics we present here, relational
models are much simpler, and as a consequence they can model richer
collections of effects than what we consider here.

\paragraph*{Step-indexing models}
The key property of full ground storage is that, when initialising a
new cell, all the information we need in order to determine its value
is in the target world. In contrast, when references can hold
functions or thunks, we also need to store the way such higher-order
values will behave in future worlds, leading to a potential cyclicity
in the semantics. Step-indexing models introduce a well-founded
hierarchy on worlds that breaks this circularity. Step-indexing
semantics to general references is given
syntactically~\cite{ahmed:thesis,
  birkedal-et-al:step-indexed-kripke-models-over-recursive-worlds,dreyer-neis-birkedal:the-impact-of-higher-order-state-and-control-effects-on-local-relational-reasoning},
or
relationally~\cite{birkedal-stovring-thamsborg:realisability-semantics-of-parametric-polymorphism-general-references-and-recursive-types,birkedal-et-al:step-indexed-kripke-models-over-recursive-worlds}. We
hope that synthesising our technique with step-indexing models or
other recursive-domain techniques would allow us to resolve the
circularity in possible worlds and extend our model structure to
general references.

\paragraph*{Games models}
Game semantics is especially well-suited for modelling local state in
the presence of higher-order functions, and full ground state is no
exception~\cite{murawski-tzevelekos:algorithmic-games-for-full-ground-references}. In
game semantics, program terms denote \emph{strategies}: dialogues
between a Player --- the program --- and an Opponent --- its
environment. In such models the semantics of the heap is implicit
(though, cf.~nominal game semantics~\cite{tzevelekos:thesis}), and
manifest in the abilities of the Opponent. In contrast, in a
sets-with-structure semantics, the heaps are fully explicit. In
particular, it might be difficult to semantically decompose a game
semantics into a monad over a bi-cartesian closed category for
modelling pure languages.

To keep the discussion precise, we give two (general) points of
comparison between the two kinds of semantics.  The
sets-with-structure semantics we present here does not validate some
equations at higher types that a game semantics would normally
validate, for example:
\[
\begin{array}{@{}l@{}l}
\infer
&
\sfun\_{\s1}\strue
\\\equiv\
\islet \cx{\isnew \strue}
{}&\sfun\_{\s1}
    \sget\cx
\end{array}
\]
The reason is that the semantics allows us to
inspect, for example, whether a value depends
on references~\cite{pitts-stark:whats-new,staton:completeness-for-algebraic-theories-of-local-state}.

In contrast, a sets-with-structure semantics makes it easier to
exploit that some types are uninhabited, for example, to prove that in
a total call-by-value language:
\[
\cx : \s1 \to \s0 \infer \strue \ \equiv\  \sfalse : \boolty
\]
where $\s0$ is the empty type.  There are currently no call-by-value
game-semantics for a total language that validate this equation. It
might be possible to develop such a model, but it would be a less
natural game semantics, as these have partiality wired in.

\paragraph*{Parametric models}
Using the semantic machinery needed to interpret parametric
polymorphism, Reddy and
Yang~\cite{reddy-yang:correctness-of-data-representations-involving-heap-data-structures}
give semantics to full ground storage. Each type denotes a functor
that has at every world, in addition to the set of values at that
world, a relation between those values. These facilitate semantic
universal and existential quantification over worlds both at the level
of types and terms.  This model also makes use of dangling
pointers. We hope further work would clarify how the extra semantic
structure in such parametric models relates to our models.

\paragraph*{Contribution}
\begin{enumerate}
\item We give a new monad for full ground storage over the category of
  functors from worlds and their morphisms into sets and functions.
\item We identify the collection of heaps as a functor from worlds and
  initialisations.
\item We decompose our full ground storage monad into a global
  state transformer applied to a monad for encapsulation.
\item We evaluate the monad in three ways:
  \begin{enumerate}
  \item We prove the full ground storage monad satisfies the effect
    masking property.
  \item We use the monad to give adequate denotational semantics to a
    total call-by-value calculus for full ground storage.
\item We show this model satisfies the usual equations for ground
  storage.
  \end{enumerate}
\end{enumerate}

The rest of the paper is structured as follows. \secref{full ground
  storage} defines full ground storage through
the syntax and operational semantics of a calculus, and highlights
where the semantic structure we expose appears in the operational
account. \secref{preliminaries} reviews the category-theoretic
background and concepts we need for our development. \secref{worlds}
defines the category of worlds and the category of
initialisations. \secref{monad} gives an explicit formula for the full
ground references monad. \secref{hiding} decomposes the monad into a
state transformed monad for encapsulation, and uses this decomposition
to analyse both the encapsulation monad and the full ground storage
monad. \secref{adequacy} returns to the calculus of \secref{full
  ground storage}, uses our monad to give denotational semantics to
this calculus, and uses its adequacy to validate that the program
equations for (ordinary) ground state carry over to the full ground
setting. \secref{conclusion} concludes.


\section{Full ground storage}\seclabel{full ground storage}
Our formalism consists of two parts. First, we fix the description of
the storable data structures in a well-founded way. We then define the
syntax, type system, and semantics of programs that manipulate data
structures involving those types.

The first component is a (typically countable) set $\cellnames$ whose elements
$\Cll$ are called \emph{cell sorts}. Given $\cellnames$, we
define the set $\grounds^{\cellnames}$ of \emph{full ground types}
$\gty$ given inductively by
\[
\gty \gdefinedby \s0 \gor \gty_1 \s+ \gty_2 \gor \s1 \gor \gty_1 \s* \gty_2 \gor \sref \Cll
\]
We omit the superscript in $\grounds^\cellnames$, and other
superscripts and subscripts, wherever possible.
The second component is a function $\typeof : \cellnames \to \grounds$
assigning to each sort its \emph{content type}: the type
of values stored in cells of this sort.

\begin{example}\examplelabel{ex1}
  To capture the example from the introduction, choose $\cellnames
  \definedby \set{\linkedlist, \cell, \data}$ and set $\typeof \Cll$,
  $\Cll \in \cellnames$, as in the introduction.
\end{example}

A \emph{full ground storage signature} is a pair $\cellsig =
\pair{\cellnames^{\cellsig}}{\typeof\mspace{-5mu}^{\cellsig}}$ of such a
set and a content type assignment for it. We can view a full ground
storage signature as an abstract description of a sequence of
top-level data declarations. Fix such a signature $\cellsig$ for the remainder of
this manuscript.

\subsection{Syntax}
\figref{syntax} presents the $\lamref$-calculus, our subject of
study. We let $\x$ range over a countable set of \emph{identifiers},
and $\ell$ range over a countably infinite set $\locs$ of
\emph{locations}. The occurrences of $\x$'s in the following
constructs are binding: function abstraction, non-empty pattern matching,
and allocation.  It is a standard call-by-value Church-style,
higher-order calculus. We \shade{\text{shade}} the parts specific to
references.

\begin{figure}
\centering
\input{cbv-syntax.tex}
\caption{The types and syntax of $\lamref$}\figlabel{syntax}
\end{figure}

We include reference literals $\ell$ which will inhabit the reference
types $\sref \Cll$. The core of high-level languages like ML does not
usually have global memory locations. However, in our core calculus we
will include values for references for two reasons. First, having
values for references makes the operational semantics straightforward
and similar in appearance to the natural global state
semantics. Second, the resulting calculus enables us to present some
program equivalences involving distinct memory locations without
introducing additional type constructors. The decision to include
memory locations in the base language is common practice in
operational and denotational semantics for local
state~\cite{benton-leperchey:relational-reasoning-in-a-nominal-semantics-for-storage,bohr-birkedal:relational-reasoning-for-recursive-types-and-references,stark:thesis,levy:global-state-considered-helpful}.

The assignment and dereferencing constructs are standard. We use a
non-standard allocation
operation~\cite{levy:global-state-considered-helpful} which allows the
simultaneous initialisation of a cyclic structure. Each of the
initialisation values $\sv_i$ has access to each of the other newly
allocated references $\x_1, \ldots, \x_n$. We require the
initialisation data to be given as values. Allowing computation at
this point would be unsound, as an arbitrary computation may try to
dereference the yet-uninitialised locations.

We will make use of the following syntactic sugar:
\[
\begin{array}{lllll}
\slet \x\ty\st{\st'}  &\equiv& (\sfun \x\ty{\st'})\ \st \\
\ssnew \Cll\st        &\equiv&
  \slet {\x}{\typeof \Cll}\st{\\&&\snew{\newbind\y\Cll\x}\y}
\\
\sseq{\st_1, \ldots, \st_n} & \equiv&
\spair{\st_1}{\spair{\cdots}{\st_n}}
\\
\ty_1\s*\cdots\s*\ty_n
&\equiv&
\ty_1\s*(\cdots\s*\ty_n)
\end{array}
\]
When it is clear from the context, we omit type annotations.  With
these conventions in place, the examples in the introduction are
special cases of full ground storage.

\begin{example-}We allocate a cyclic list:
    \vspace{-.5cm}

{
    \newcommand\Payload{\mathtt{payload}}
    \newcommand\Lst{\mathtt{cyclic\_list}}
    \newcommand\Head{\mathtt{head}}

  \hfill
  \begin{tikzpicture}[list/.style={rectangle split, rectangle split parts=2,
    draw, rectangle split horizontal, rectangle split draw splits=false}, >=stealth, start chain, node distance=1mm]

  \node[on chain] (B) {\framebox{$42$}};
  \node[on chain] (C) {\framebox{$\sinj[]2\hphantom{\bullet}$}};
  \node[list, on chain] (A) {\,$\mathllap{(}$\hspace{10pt}\nodepart{two}\hspace{-3pt},\hspace{4.5pt}$\mathrlap{)}$\,};
  \draw[*->] let \p1 = (C.center), \p2 = (A) in (\x1 + 7, \y1) -- (A);
  \draw[*->] let \p1 = (A.two), \p2 = (A.center), \p3 = (C.west) in
  (\x1,\y2) -- (\x1 + 15, \y2) -- (\x1 + 15, \y2 - 15) -- (\x3 - 2.5, \y2 - 15)
            -- (\x3 - 2.5, \y2) -- (\x3+4, \y3);
  \draw[*->] let \p1 = (A.one), \p2 = (B) in (\x1+3,\y1 ) -- (\x1 + 3, \y1 + 15) -- (\x2, \y1 + 15) -- (\x2, \y2 +8);
  \draw[*->] let \p1 = (B.south), \p2 = (C.south west) in
  (\x1, \y1-2) -- (\x2+7, \y1-2) -- (\x2+7, \y1+2);
  \node at  ($(B.south west) + (.25,-.1)$) {$\mathllap{\Lst}$};
  \end{tikzpicture}

\vspace{-1cm}
    \[
      \begin{array}{*1{@{}l}}
         \snew {\\\ \begin{array}[t]{*4{@{}l}}
                  {}&
                      \newbind[{&}]{\Payload&}{\data  }{42}\newsep \\
                  {}&
                      \newbind[{&}]{\Lst&}{\linkedlist}{\sinj[]2{\Head}}\newsep \\
                  {}&
                      \newbind[{&}]{\Head&}{\cell     }{\spair\Payload\Lst}
                \end{array}\\
         }{\Lst}
      \end{array}
    \]
}
\\[-2\baselineskip]\qed
\end{example-}

Our type-system needs to associate a sort to each
location literal $\ell$ the program may use. A \emph{heap layout} $\w$
is a partial function with finite support $\w : \locs \pfinto
\cellnames$. We write $\set{\ell_1 : \Cll_1, \ldots, \ell_n : \Cll_n}$
for the heap layout whose support is $\ord\w \definedby \set{\ell_1, \ldots, \ell_n}$
defined by $\w(\ell_i) = \Cll_i$.  When $\w(\ell) = \Cll$, we write
$(\ell : \Cll) \in \w$. We write $\w \leq \w'$ when $\w'$ extends
$\w$. Layout extension is thus a partial order. Heap layouts are a
standard abstraction and appear under different names:
state-types~\cite{benton-leperchey:relational-reasoning-in-a-nominal-semantics-for-storage,bohr-birkedal:relational-reasoning-for-recursive-types-and-references},
and location
worlds~\cite{reddy-yang:correctness-of-data-representations-involving-heap-data-structures}.

\begin{figure}
  \centering
  \input{cbv-type-system.tex}
  \caption{The inductive definition of the typing relation of
    $\lamref$}\figlabel{type system}
\end{figure}

Typing contexts $\Tys$ are partial functions with finite support from
the set of identifiers to the set of types. We use the list-like
notation $\Tys, \x : \ty$ for the extension of $\Tys$ by the
assignment $\x \mapsto \ty$, and the membership-like notation $(\x :
\ty) \in \Tys$ to state that $\Tys(\x) = \ty$. The type system is
given in \figref{type system} via an inductively defined quaternary
relation $\Tys \infer_{\w} \st : \ty$ between contexts
$\Tys$, layouts $\w$, terms $\st$, and types $\ty$.

Location literals $\ell$ are limited to the locations in the layout
$\w$, which does not change throughout the typing derivation. To
assign, the type of the assigned value needs to match the sort of the
reference, and similarly the type of the dereferenced value matches
that of the reference. Finally, for allocation, the initialisation
values may refer to each of the newly allocated references, as does
the remainder of the computation.

By construction, every typeable term has a unique type in a given
context.

A \emph{value substitution} $\e$ is a partial function with
finite support from the set of identifiers to values. Defining capture
avoiding substitution $\subst\st\e$ and proving the substitution lemma
is standard and straightforward. In the sequel we will hand-wave
around the standard issues with $\alpha$-equivalence, and omit the
standard freshness conditions on bound variables.

Finally, the type
system is monotone with respect to layout extension: $\Tys \infer_{\w}
\st: \ty$, and $\w' \geq \w$ implies $\Tys \infer_{\w'} \st : \ty$. In
particular, if we define $\ty\w$ to be the set of closed values of
type $\ty$ assuming layout $\w$, $\ty\w$ is functorial in $\w$ in the
following sense: for every $\w \leq \w'$, $\ty\w \subset \ty\w'$.

\subsection{Operational semantics}
We present a big-step operational semantics. We expect a small-step
semantics or stack-machine semantics to be similarly
straightforward.

An \emph{untyped heap} $\heap$ is a partial function with finite
support $\ord\w_{\heap}$ from $\locs$ to values.  A \emph{typed heap}
$\heap$ consists of a pair of a layout $\w_{\heap}$, and a function
from the set of locations in $\w_{\heap}$ to values that assigns to
every $(\ell : \Cll) \in \w_{\heap}$ a well-typed closed value:
$\infer_{\w_{\heap}}\heap(\ell) : \typeof \Cll$. We denote the set of
heaps with layout $\w$ by $\heaps\w$. For every layout $\w$, an
untyped heap can be turned into a typed heap from $\heaps\w$ in at
most one way. Note that $\heaps\w$ is not functorial with respect to
layout extension: there is no obvious way to
turn an arbitrary heap in $\heaps\w$ into a heap in $\heaps\w'$ for
every $\w' \geq \w$.

\begin{figure}
  \centering
  \input{cbv-op-sem.tex}
  \caption{The operational semantics of $\lamref$}\figlabel{op sem}
\end{figure}

A \emph{configuration} is a pair $\pair\st\heap$ consisting of a term
$\st$ and an untyped heap $\heap$. A \emph{terminal} configuration is
one whose term is a value.  We say that a sequence of locations
$\ell_1, \ldots, \ell_n$ is \emph{fresh} for layout $\w$, and write
$\fresh\w{\ell_1, \ldots, \ell_n}$, when the locations are pairwise
distinct and collectively disjoint from $\w$'s support. \figref{op
  sem} defines the evaluation relation
$\pair\st\heap\bigstep\pair\sv{\heap'}$ between configurations and
terminal configurations.

Locations, as values, are fully evaluated. Assignment updates the
heap, and dereferencing retrieves the appropriate value from the
heap. The rule for allocation requires the newly allocated locations
to be fresh for the current heap, and then extends the heap with the
given initialisation values with the new locations substituted in. It
then carries the execution in the body of the allocation with those
new locations substituted in. As the only requirement of the new
location is to be fresh, this semantics is not deterministic,
and a phrase might evaluate to several different configurations with
different heap layouts.

We prove Felleisen-Wright soundness for $\lamref$.

\begin{theorem}[preservation]\theoremlabel{preservation}
  Evaluation preserves typeability: for every well-typed closed term
  $\infer_{\w} \st:\ty$ and for every $\w_1 \geq \w$ and $\heap_1 \in
  \heaps\w_1$, if $\pair\st{\heap_1} \bigstep \pair\sv{\heap_2}$, then there
  is some $\w_2 \geq \w_1$ such that $\infer_{\w_2} \sv:\ty$ and
  $\heap_2 \in \heaps\w_2$.
\end{theorem}

The proof is by straightforward induction on the evaluation relation
after strengthening the induction hypothesis to closed substitutions
in open terms.

\begin{theorem}[totality]\theoremlabel{totality}
  All well-typed closed programs fully evaluate: for every
  $\infer_{\w} \st:\ty$, $\w_1 \geq \w$ and $\heap_1\in\heaps\w_1$
  there exist some $\w_2 \geq \w_1$, $\infer_{\w_2} \sv: \ty$, and
  $\heap_2 \in \heaps\w_2$ such that
  $\pair\st{\heap_1}\bigstep\pair\sv{\heap_2}$.
\end{theorem}

The proof is standard using Tait's
method~\cite{tait-intensional-interpretations-of-functionals-of-finite-type-i}
and Kripke logical predicates, and a $\w$-indexed predicate on
$\heaps\w$ that is \emph{not} Kripke, as $\heaps\w$ is not functorial
in layout extensions.

We focus on the following aspects from the allocation case in the
course of this proof. In that case, the allocation construct is typed
with respect to heap layout $\w$, but operates on a heap with an
extended layout $\w \leq \w'$.  We can find fresh locations $\ell_1,
\ldots, \ell_n$, and then form the following square of layout
extensions
\[
\begin{array}{*2{l@{}}c}
  \w &{}\leq \w&\+\set{\ell_1:\Cll_1, \ldots, \ell_n:\Cll_n} \\
  \rotatebox[origin=c]{-90}{$\leq$} &&\rotatebox[origin=c]{-90}{$\leq$}\\
  \w'&{}\leq\w'&\+\set{\ell_1:\Cll_1, \ldots, \ell_n:\Cll_n}
\end{array}
\]
where the operation $\+$ denotes layout extension by the given
locations and sorts.  The initialisation data in the allocation
construct is given for the top extension. The crucial step in the
proof in this case is that we can transform this initialisation data
into initialisation data for the extension in the bottom row. Applying
this transformed initialisation data on the given heap is precisely
the functorial action of the collection of heaps.

\subsection{Observational equivalence}\subseclabel{observational equivalence}
To define observational equivalence, we need a few more technical
definitions. Let $\Tys$, $\Tys'$ be two typing contexts. We say that
$\Tys'$ \emph{extends} $\Tys$, and write $\Tys' \geq \Tys$ when
$\Tys'$ extends $\Tys$ as a function from identifier names to types.
In the sequel we write $\judge{\st, \st'}\ty$ to indicate that the
quintuple $\seq{\Tys,\w,\st,\st',\ty}$ belongs to the quinary relation
given by the conjunction $\judge\st\ty$ and $\judge{\st'}\ty$.

Consider any two terms $\judge {\st,\st'}\ty$. We define the \emph{set
  of contexts plugged with ${\judge {\st, \st'}\ty}$}, which we denote
by $\Plugged\Tys\w\st{\st'}\ty$, as the smallest quinary relation
jointly compatible with the typing rules that contains the quintuples
$\seq{\Tys', \w', \st, \st', \ty}$ for every  \( \Tys'\infer_{\w'}\st,{\st'}:\ty \),   $\Tys' \geq
\Tys$, and $\w' \geq \w$.

We say that two terms $\judge{\st_1,\st_2}\ty$ are \emph{observationally
  equivalent} when for all closed boolean plugged contexts
$\infer_{\w'}{\st'_1,\st'_2}:\boolty \in \Plugged\Tys\w{\st_1}{\st_2}\ty$,
heaps $\heap' \in \heaps\w'$ and boolean values
$\infer\sv:\boolty $, we have:
\begin{mathpar}
  \exists{\heap_1}(\pair{\st'_1}{\heap'} \bigstep \pair{\sv}{\heap_1})

  \iff

  \exists{\heap_2}(\pair{\st'_2}{\heap'} \bigstep \pair{\sv}{\heap_2})
\end{mathpar}
Note that by the Preservation \theoremref{preservation}, if such heaps
$\heap_i$ exist, they can be typed.


\section{Preliminaries}\seclabel{preliminaries}
We assume familiarity with categories, functors, and natural
transformations. We denote by $\Set$ the category of sets and
functions. Let $\scat$ be a small category.  We denote the category of
functors $X, Y : \scat \to \Set$ and natural transformations between
them by $[\scat, \Set]$. As we will interpret types in such a
category, we recall its bi-cartesian closed structure: its sums and
products are given component-wise, and the exponential is given by an
end formula (see below). We assume familiarity with ends and coends
over $\Set$, which we will use in the explicit description of the full
ground storage monad. To fix terminology and notation, given a mixed
variance functor $P : \opposite\scat\times\scat \to \Set$, we denote
its end and its ending wedge as follows, for all $\w' \in \scat$:
\[
\projection_{\w'} : \cocoend{\w\in\scat} P(\w,\w) \to P(\w', \w')
\]
We denote its coend and its coending wedge as follows:
\[
q_{\w'} : P(\w', \w') \to \coend{\w\in\scat} P(\w, \w)
\]

Consider any functor $\p : \init \to \worlds$  between two small
categories. Precomposition with $\p$ induces a functor ${\p^* :
[\worlds, \Set] \to [\init, \Set]}$. By generalities, this
functor has a right adjoint $\p_* : [\init, \Set] \to [\worlds,
  \Set]$ called the \emph{right Kan extension along $\p$}, which we will
use to establish the existence of the monoidal strength structure for
our monad. Let $\w \in \worlds$ be an object in $\worlds$. We will use
the following specific kind of comma category in our ends and
coends. The comma category $\w\comma\p$ has as objects pairs
$\pair{e}{\h}$ where $e \in \init$ is an object and $\h : \w \to \p e$
is a morphism in $\worlds$. Its morphisms $\ih : \pair{e_1}{\h_1} \to
\pair{e_2}{\h_2}$ are morphisms $\ih : e_1 \to e_2$ in $\init$ such
that $\p\ih \compose \h_1 = \h_2$. As we assumed $\init$ and $\worlds$
are small, so is $\w\comma\p$. Precomposition with the projection
functor ${\projection : (\w\comma\p) \to \init}$
turns every mixed variance functor $P : \opposite\init\times\init \to
\Set$ into a mixed variance functor $\commatise P :
\opposite{(\w\comma\p)}\times(\w\comma\p) \to \Set$. There is then a
canonical isomorphism:
\[
\cocoend{\w\comma\p}\commatise P
=
\cocoend{\pair{e}{\h} \in \w\comma\p} P(e,e)
\isomorphic
\cocoend{e\in\init}(P(e,e))^{\worlds(\w, \p e)}
\]
Also recall the end formula for exponentials in $[\worlds, \Set]$:
\[
X^Y\w \isomorphic \cocoend{\w' \in \worlds} (X\w')^{\worlds(\w, \w')\times Y\w'}
\isomorphic
\cocoend*{\h : \w \to \w'\in \w\comma\id[\worlds]} \quad(X\w')^{(Y\w')}
\]

To describe the tensorial strengths for our monads, we assume
familiarity with symmetric monoidal closed categories, though
concretely we will only use a cartesian structure. The following
concept will simplify the presentation of the strengths. Let $\moncat
\vals = \seq{\vals, \*, \I, \assoc, \lunit, \runit}$ be a symmetric
monoidal closed category. A \emph{monoidal $\moncat\vals$-action
  category}, also known as a \emph{$\moncat\vals$-actegory}, is a tuple
$\monact\pconfs = \seq{\pconfs, \., \compact, \trivact}$ where:
\begin{itemize}
\item $\. : \vals \times \pconfs \to \pconfs$ is a two-argument functor;
\item $\compact_{X,Y, A} : (X\*Y)\.A \to X\.(Y\.A)$ is a natural isomorphism; and
\item $\trivact_{A} : \I\.A \to A$ is a natural isomorphism
\end{itemize}
subject to the following coherence axioms:
\begin{gather*}
  \begin{aligned}
    \compact_{X, Y, Z\.A}&\compose\compact_{X\*Y, Z, A}
    \\&=
    \id[X]\.\compact_{Y, Z, A}\compose\compact_{X, Y\*Z, A}\compose\assoc[X,
      Y, Z]\.\id[A]
  \end{aligned}
  \\
  \id[X]\.\trivact_{A}\compose\compact_{X, \I, A} = \runit[X]\.\id[A]
\end{gather*}
We say that a $\moncat\vals$-actegory $\monact\pconfs$ is
\emph{bi-closed} if the following two right adjoints exist:
\begin{itemize}
\item $\placeholder \,\. A \leftadjointto A \acthom \placeholder : \pconfs
  \to \vals$ for every $A \in \pconfs$
\item $X\.\placeholder \leftadjointto X \powers \placeholder : \pconfs
  \to \pconfs\ $ for every $X \in \vals$.
\end{itemize}
We denote the natural bijection of the first adjunction by
$\curry^{\acthom}_{X, A, B} : \vals(X\.A, B) \to \pconfs(X, A\acthom
B)$, its inverse by $\uncurry^{\acthom}$, and its counit by
$\eval^{\acthom}$, and similarly for $\powers$.

As the name suggests, a $\moncat\vals$-actegory is a
$\moncat\vals$-category ($\moncat\vals$-enriched category) with the
structure permuted:
\begin{theorem*}[\!\cite{janelidze-kelly:a-note-on-actions-of-a-monoidal-cat,gordon-power:enrichment-through-variation}]
  These data are canonically isomorphic:
  \begin{itemize}
  \item a bi-closed $\moncat\vals$-actegory $\monact\pconfs$; and
  \item a $\moncat\vals$-category $\monact\pconfs$ with powers and
    copowers.
  \end{itemize}
  and moreover, the adjoints $\acthom$ and $\powers$ enrich.
\end{theorem*}
Up to isomorphism, the enrichment is given by the adjoint $\acthom$,
the copowers by the monoidal action $\.$, and the powers by the
adjoint $\powers$. The relevance of $\moncat\vals$-actegories to our
situation is that the monoidal action in our setting is much simpler
than the enrichment. Working with monoidal actions thus simplifies
many calculations.

Finally, we assume familiarity with monads, formulated as Kleisli
triples. Let $\structure T = \triple T\mreturn\bind$ be a monad over a
$\moncat\vals$-actegory $\monact\pconfs$. A \emph{tensorial strength}
for $\moncat T$ is a natural transformation $\strength : X\.TA \to
T(X\.A)$ satisfying the following coherence axioms:
\begin{gather*}
  T\trivact_{A}\compose\strength_{\I, A} = \trivact_{TA}\\
  \strength_{X, Y\.A}\compose\id[X]\.\strength_{Y, A}\compose \compact_{X, Y, TA}
  =
  T\compact_{X, Y, A}\compose\strength_{X\*Y, A}\\
  \strength_{X, A}\compose\id[X]\.\mreturn_{A} = \mreturn_{X\.A} \\
  (\bind\!\id[T(X\.A)])\compose T\strength_{X, A}\compose\strength_{X, TA}
  =
  \strength_{X, A}\compose\id[X]\.\!\bind\!\id[TA]
\end{gather*}
Let $\structure T$ and $\structure S$ be two such strong monads. We
say that a monad morphism $m : \structure T \to \structure S$ is
\emph{strong} when for all $X \in \worlds, A \in \pconfs$:
$m_{X\.A}\compose\strength_{X,A} = \strength_{X, SA}\compose
\id[X]\.m$.
When the actegory is bi-closed, the data for a strong monad and the
data for a $\moncat\vals$-monad are canonically
isomorphic~(cf.~\cite{kock:strong-functors-and-monoidal-monads}).

We are now ready to describe the state monad transformer that will
give our monad for reference cells, as a straightforward consequence
of transforming a monad across the adjunction
$\placeholder\. A\leftadjointto A\acthom\placeholder$
(see also e.g.~\cite{moegelberg-staton:linear-state}):
\begin{proposition}\propositionlabel{global state monad transformer}
Let $\moncat\vals$ be a symmetric monoidal closed category and
$\monact\pconfs$ be a bi-closed $\moncat\vals$-actegory. Let $A$ be an
object in $\pconfs$.
For every strong monad $P$ over $\monact\pconfs$, we have a strong
monad $\structure T = \set{T, \mreturn, \bind, \strength}$ over
$\moncat\vals$ given by:
\begin{mathpar}
    T A \definedby A \acthom P(\placeholder \. A)

    \mreturn^T_X \definedby \curry^{\acthom}_{X, A, P(A\.A)}\parent{\mreturn^P_{X\.A}}

    \bind^T_{X,Y} f
    \definedby
    A\acthom (\bind_{X\.A, Y\.A}^P\uncurry_{X, A, P(Y\.A)}^{\acthom}f)

    \begin{aligned}
    \strength^T_{X, Y}& \definedby \curry^{\acthom}_{X\*TY, A, P((X\*Y)\.A)}
    \\
    &\mspace{-30mu}(
    P\inv\compact_{X, Y, A}\compose\strength^P_{X, Y\.A}\compose(\id[X]\.\eval^{\acthom}_{A, P(Y\.A)})\compose\compact_{X, TY, A})
    \end{aligned}
\end{mathpar}
    and for every strong monad
    morphism $m : \structure P_1 \to \structure P_2$ we have a strong monad
    morphism $m^T : T_1 \to T_2$ between the corresponding monads,
    given by:
    \[
    m^T_X \definedby A\acthom(m_{X\.A})
    \]
    \cdiagram{situation-01}
\end{proposition}


\section{Worlds and initialisations}\seclabel{worlds}
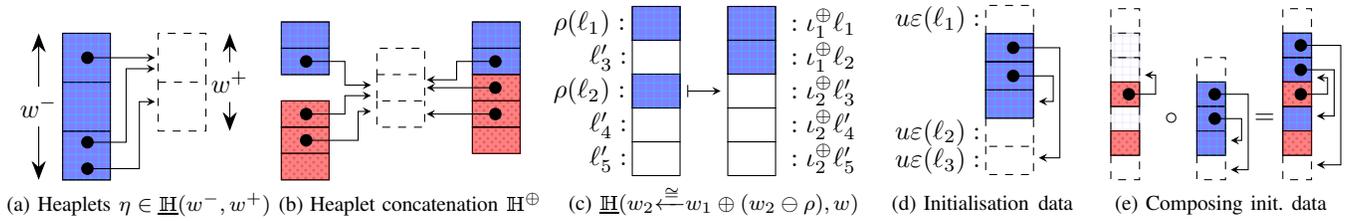
\begin{figure*}
\hfil%
\subfloat[Heaplets $\store \in \prostores(\w^-, \w^+)$]{
\begin{tikzpicture}[x=.25in,y=0.65cm, >=stealth]
  \foreach \x in {1,...,3}
  {
    \node[inner sep=1pt] (BL\x) at (0,-\x)   {};
    \node[inner sep=1pt] (UR\x) at (1,-\x+1) {};
    \node (MM\x) at ($(BL\x)!0.5!(UR\x)$) {};
    \draw (BL\x) rectangle (UR\x);

  }

  \foreach \i in {5}
  {
    \DashedHeapCell{(BL\i)}{(UR\i)}{(2, +4-\i)}
  }
  \foreach \i in {6}
  {
    \DashedCellDownwards{(BL\i)}{(UR\i)}{(2, +4-\i)}
  }

  \patternA{(BL1)}{(UR1)}
  \patternA{(BL2)}{(UR2)}
  \patternA{(BL3)}{(UR3)}

  \draw[*->] let \p1 = ($(BL1)!0.5!(UR1)$),
                 \p2 = ($(BL5)!0.6!(UR5)$)
             in (\x1-2, \y1) -- (\x1 + 10 + 2*5, \y1)
                             -- (\x1 + 10 + 2*5, \y1)
                             -- (\x2 - 12, \y1);

  \draw[*->] let \p1 = ($(BL3)!0.5!(UR3)$),
                 \p2 = ($(BL5)!0.6!(UR5)$)
             in (\x1-2, \y1+5) -- (\x1 + 10 + 1*5, \y1+5)
                             -- (\x1 + 10 + 1*5, \y2-6)
                             -- (\x2 - 12, \y2-6);

  \draw[*->] let \p1 = ($(BL3)!0.5!(UR3)$),
                 \p2 = ($(BL6)!0.6!(UR6)$)
             in (\x1-2, \y1-5) -- (\x1 + 10 + 2*5, \y1-5)
                             -- (\x1 + 10 + 2*5, \y2)
                             -- (\x2 - 12, \y2);

  \node (lab) at ($(BL1)!0.5!(BL3)+(-.5, .5)$)  {$\w^-$};
  \draw[-{Stealth[scale=1.5]}]
       (lab) -- ($(BL1) + (-.5,1)$) ;
  \draw[-{Stealth[scale=1.5]}]
       (lab) -- ($(BL3) + (-.5,0)$);

  \node (lab) at ($(BL5)!0.5!(BL6)+(1.5, .5)$)  {$\w^+$};
  \draw[-{Stealth[scale=1.5]}]
       (lab) -- ($(BL5) + (1.5,1)$) ;
  \draw[-{Stealth[scale=1.5]}]
  (lab) -- ($(BL6) + (1.5,0)$);

\end{tikzpicture}
\subfiglabel{heaplet def}}%
\hfil%
\subfloat[Heaplet concatenation $\storesmon$]{%
\begin{tikzpicture}[x=.25in,y=0.35cm,>=stealth]
  \foreach \x in {0,1}
  {
    \node[inner sep=1pt] (BL\x) at (0,-\x)   {};
    \node[inner sep=1pt] (UR\x) at (1,-\x+1) {};
    \node (MM\x) at ($(BL\x)!0.5!(UR\x)$) {};
    \draw (BL\x) rectangle (UR\x);

    \patternA{(BL\x)}{(UR\x)};
  }

  \foreach \x in {2,3,4}
  {
    \node[inner sep=1pt] (BL\x) at (0,-1-\x)   {};
    \node[inner sep=1pt] (UR\x) at (1,-1-\x+1) {};
    \node (MM\x) at ($(BL\x)!0.5!(UR\x)$) {};
    \draw (BL\x) rectangle (UR\x);
    \patternB{(BL\x)}{(UR\x)};
  }

  \node[inner sep=1pt] (MBL0) at (2,-1-0)   {};
  \node[inner sep=1pt] (MUR0) at (3,-1-0+1) {};
  \node (MMM0) at ($(MBL0)!0.5!(MUR0)$) {};
  \draw[dashed] (MBL0) rectangle (MUR0);

  \foreach \i in {1,2}
  {
    \DashedCellDownwards{(MBL\i)}{(MUR\i)}{(2, -1-\i)};
  }

  \draw[*->] let \p1 = ($(BL1)!0.5!(UR1)$),
                 \p2 = ($(MBL1)!0.6!(MUR1)+(-1,0)$)
             in (\x1-2, \y1) -- (\x1 + 5 + 2*5, \y1)
                             -- (\x1 + 5 + 2*5, \y2)
                             -- (\x2 +5, \y2);

  \draw[*->] let \p1 = ($(BL2)!0.5!(UR2)$),
                 \p2 = ($(MBL1)!0.6!(MUR1)+(-1,0)$)
             in (\x1-2, \y1) -- (\x1 + 5 + 2*5, \y1)
                             -- (\x1 + 5 + 2*5, \y2-4)
                             -- (\x2 + 5, \y2-4);

  \draw[*->] let \p1 = ($(BL3)!0.5!(UR3)$),
                 \p2 = ($(MBL2)!0.6!(MUR2)+(-1,0)$)
             in (\x1-2, \y1) -- (\x1 + 5 + 3*5, \y1)
                             -- (\x1 + 5 + 3*5, \y2)
                             -- (\x2 + 5, \y2);

  \foreach \x in {0,1}
  {
    \node[inner sep=1pt] (BL\x) at (4,-\x)   {};
    \node[inner sep=1pt] (UR\x) at (5,-\x+1) {};
    \node (MM\x) at ($(BL\x)!0.5!(UR\x)$) {};
    \draw (BL\x) rectangle (UR\x);
    \patternA{(BL\x)}{(UR\x)};
  }

  \foreach \x in {2,3,4}
  {
    \node[inner sep=1pt] (BL\x) at (4,-\x)   {};
    \node[inner sep=1pt] (UR\x) at (5,-\x+1) {};
    \node (MM\x) at ($(BL\x)!0.5!(UR\x)$) {};
    \draw (BL\x) rectangle (UR\x);
    \patternB{(BL\x)}{(UR\x)};
  }

  \draw[*->] let \p1 = ($(BL1)!0.5!(UR1)$),
                 \p2 = ($(MBL1)!0.6!(MUR1)$)
             in (\x1+2, \y1) -- (\x1 - 5 - 2*5, \y1)
                             -- (\x1 - 5 - 2*5, \y2+2)
                             -- (\x2 + 8, \y2+2);

  \draw[*->] let \p1 = ($(BL2)!0.5!(UR2)$),
                 \p2 = ($(MBL1)!0.6!(MUR1)$)
             in (\x1+2, \y1) -- (\x2 +8, \y1);

  \draw[*->] let \p1 = ($(BL3)!0.5!(UR3)$),
                 \p2 = ($(MBL2)!0.6!(MUR2)$)
             in (\x1+2, \y1) -- (\x2 + 8 , \y1);

\end{tikzpicture}
\subfiglabel{heaplet concat}}%
\hfil%
\subfloat[$\prostores(\w_2 \smash{\xfrom{\isomorphic}} \w_1\+(\w_2\o-\h), \w)$]{%
\begin{tikzpicture}[x=.25in,y=0.45cm,>=stealth]

  \foreach \x in {0,1,2,3,4}
  {
    \node[inner sep=1pt] (BL\x) at (0,-\x)   {};
    \node[inner sep=1pt] (UR\x) at (1,-\x+1) {};
    \node (MM\x) at ($(BL\x)!0.5!(UR\x)$) {};
    \draw (BL\x) rectangle (UR\x);
  }

  \foreach \x/\y in {0/1,2/2}
  {
    \node at ($(MM\x)-(1.4,0)$) {$\h(\ell_\y):$};
    \patternA{(BL\x)}{(UR\x)}
  }

  \foreach \x/\y in {1/3,3/4, 4/5}
  {
    \node at ($(MM\x)-(1,0)$) {$\ell'_\y:$};
  }

  \foreach \x/\y in {0/1,1/2}
  {
    \node[inner sep=1pt] (RBL\x) at (2,-\x)   {};
    \node[inner sep=1pt] (RUR\x) at (3,-\x+1) {};
    \node (MM\x) at ($(RBL\x)!0.5!(RUR\x)$) {};
    \draw (RBL\x) rectangle (RUR\x);
    \patternA{(RBL\x)}{(RUR\x)}
    \node at ($(MM\x)+(1.5,0)$) {$:\iinj_1\ell_\y$};
  }

  \foreach \x/\y in {2/3,3/4,4/5}
  {
    \node[inner sep=1pt] (RBL\x) at (2,-\x)   {};
    \node[inner sep=1pt] (RUR\x) at (3,-\x+1) {};
    \node (RMM\x) at ($(RBL\x)!0.5!(RUR\x)$) {};
    \draw (RBL\x) rectangle (RUR\x);
    \node at ($(RMM\x)+(1.5,0)$) {$:\iinj_2\ell'_\y$};
  }

  \draw[|->] let \p1 = ($(BL2)+(1, 0)$),
                 \p2 = (RBL2)
             in ($(MM2)+(.65, 0)$) -- ($(RMM2)-(.65, 0)$);

\end{tikzpicture}
\subfiglabel{heaplet defrag}}%
\hfil%
\subfloat[Initialisation data]{%
\begin{tikzpicture}[x=.25in,y=0.375cm,>=stealth]
  \node[inner sep=1pt] (BL0) at (0,-0)   {};
  \node[inner sep=1pt] (UR0) at (1,-0+1) {};
  \node (MM0) at ($(BL0)!0.5!(UR0)$) {};
  \draw[dashed]
          let \p1 = (BL0),
              \p2 = (UR0)
          in (\x1,\y1) -- (\x1, \y2) -- (\x2, \y2) -- (\x2, \y1) ;
    \node at ($(MM0) + (-1.5, 0)$) {$\p\ih(\ell_{1}):$};

  \foreach \i/\j in {4/2,5/3}
  {
    \node[inner sep=1pt] (BL\i) at (0,-\i)   {};
    \node[inner sep=1pt] (UR\i) at (1,-\i+1) {};
    \node (MM\i) at ($(BL\i)!0.5!(UR\i)$) {};
    \node at ($(MM\i) + (-1.5, 0)$) {$\p\ih(\ell_{\j}):$};
    \draw[dashed]
          let \p1 = (BL\i),
              \p2 = (UR\i)
          in (\x1, \y2) -- (\x1,\y1) -- (\x2, \y1) -- (\x2, \y2);
  }

  \foreach \x in {1,2,3}
  {
    \node[inner sep=1pt] (BL\x) at (0,-\x)   {};
    \node[inner sep=1pt] (UR\x) at (1,-\x+1) {};
    \node (MM\x) at ($(BL\x)!0.5!(UR\x)$) {};
    \draw (BL\x) rectangle (UR\x);
    \patternA{(BL\x)}{(UR\x)};
  }

  \draw[*->] let \p1 = ($(BL1)!0.5!(UR1)+(-1pt, 0)$),
                 \p2 = ($(BL5)!0.5!(UR5)+ (.5, 0) + (2pt, 1pt)$)
             in (\x1, \y1) -- (\x1 + 15pt + 2*2.5pt, \y1)
                           -- (\x1 + 15pt + 2*2.5pt, \y2)
                           -- (\x2, \y2);

  \draw[*->] let \p1 = ($(BL2)!0.5!(UR2)+(-1pt, 0)$),
                 \p2 = ($(BL3)!0.5!(UR3)+ (.5, 0) + (2pt, 1pt)$)
             in (\x1, \y1) -- (\x1 + 15pt + 1*2.5pt, \y1)
                           -- (\x1 + 15pt + 1*2.5pt, \y2)
                           -- (\x2, \y2);

\end{tikzpicture}
\subfiglabel{init data def}}%
\hfil%
\subfloat[Composing init.~data]{%
\hspace{1em}%
\begin{tikzpicture}[x=.15in,y=0.325cm, >=stealth]
  \DashedCellUpwards[(MM0)]{(BL0)}{(UR0)}{(3, 0)}

  \foreach \i in {4}
  {
    \DashedCellDownwards[(MM\i)]{(LU\i)}{(UR\i)}{(3, -\i)}
  }

  \foreach \x in {1,2,3}
  {
    \node[inner sep=1pt] (BL\x) at (3,-\x)   {};
    \node[inner sep=1pt] (UR\x) at (4,-\x+1) {};
    \node (MM\x) at ($(BL\x)!0.5!(UR\x)$) {};
    \draw (BL\x) rectangle (UR\x);
    \patternA{(BL\x)}{(UR\x)};
  }

  \draw[*->] let \p1 = ($(BL1)!0.5!(UR1)+(-1pt, 0)$),
                 \p2 = ($(BL4)!0.5!(UR3)+ (2, -.5) + (2pt, 2pt)$)
             in (\x1, \y1) -- (\x1 + 10pt + 2*2.5pt, \y1)
                           -- (\x1 + 10pt + 2*2.5pt, \y2)
                           -- (\x2, \y2);

  \draw[*->] let \p1 = ($(BL2)!0.5!(UR2)+(-1pt, 0)$),
                 \p2 = ($(BL3)!0.5!(UR3)+ (.5, 0) + (2pt, 1pt)$)
             in (\x1, \y1) -- (\x1 + 10pt + 1*2.5pt, \y1)
                           -- (\x1 + 10pt + 1*2.5pt, \y2)
                           -- (\x2, \y2);

   \foreach \i in {0, 1, 2}
   {
     \DashedCellUpwards[(MM\i)]{(BL\i)}{(UR\i)}{(0, 2-\i)}
   }

   \DashedCellMiddle[(MM4)]{(BL4)}{(UR4)}{(0, 2-4)}

   \foreach \i in {1, 2, 4}
   {
     \patternA*{(BL\i)}{(UR\i)}
   }

   \foreach \i in {3, 5}
   {
     \HeapCell[{\patternB{(BL\i)}{(UR\i)}}]{(BL\i)}{(UR\i)}{(0, 2-\i)}
   }

   \foreach \i in {6}
   {
     \DashedCellDownwards[(MM\i)]{(BL\i)}{(UR\i)}{(0, 2-\i)}
   }

   \draw[*->] let \p1 = ($(BL3)!0.5!(UR3)+(-1pt, 0)$),
                  \p2 = ($(BL2)!0.5!(UR2)+ (.5, 0) + (2pt, 1pt)$)
             in (\x1, \y1) -- (\x1 + 10pt + 1*2.5pt, \y1)
                           -- (\x1 + 10pt + 1*2.5pt, \y2-3pt)
                           -- (\x2, \y2-3pt);

   \node at ($(UR4.east)!0.5!(UR5.east) + (1, 0)$) {$\compose$};
   \node at ($(UR4.east)!0.5!(UR5.east) + (4.25, 0)$) {$=$};

   \DashedCellUpwards{(N1)}{(N2)}{(6, 2 - 0)}

   \foreach \i in {1,2, 4}
   {
     \HeapCell[{\patternA{(BL\i)}{(UR\i)}}]{(BL\i)}{(UR\i)}{(6, 2-\i)}
   }

   \foreach \i in {3, 5}
   {
     \HeapCell[{\patternB{(BL\i)}{(UR\i)}}]{(BL\i)}{(UR\i)}{(6, 2-\i)}
   }

   \foreach \i in {6}
   {
     \DashedCellDownwards[(MM\i)]{(BL\i)}{(UR\i)}{(6, 2-\i)}
   }

   \draw[*->] let \p1 = ($(BL3)!0.5!(UR3)+(-1pt, 0)$),
                  \p2 = ($(BL2)!0.5!(UR2)+ (.5, 0) + (2pt, 1pt)$)
             in (\x1, \y1) -- (\x1 + 10pt + 1*2.5pt, \y1)
                           -- (\x1 + 10pt + 1*2.5pt, \y2-4pt)
                           -- (\x2, \y2-4pt);

   \draw[*->] let \p1 = ($(BL2)!0.5!(UR2)+(-1pt, 0)$),
                  \p2 = ($(BL4)!0.5!(UR4)+ (.5, 0) + (2pt, 1pt)$)
             in (\x1, \y1) -- (\x1 + 10pt + 2*2.5pt, \y1)
                           -- (\x1 + 10pt + 2*2.5pt, \y2)
                           -- (\x2, \y2);

   \draw[*->] let \p1 = ($(BL1)!0.5!(UR1)+(-1pt, 0)$),
                  \p2 = ($(BL6)!0.5!(UR6)+ (.5, 0) + (2pt, 1pt)$)
             in (\x1, \y1) -- (\x1 + 10pt + 3*2.5pt, \y1)
                           -- (\x1 + 10pt + 3*2.5pt, \y2)
                           -- (\x2, \y2);

\end{tikzpicture}
\subfiglabel{init data composition}}
\hfil%
\caption{Heaplets, initialisations, and their operations}\figlabel{heaplets}
\end{figure*}

In \secref{full ground storage}, we arranged the terms and values of
$\lamref$ based on their heap layout assumptions, $\w$, and this
arrangement is functorial with respect to layout extension $\w \leq
\w'$. We saw that heaps also divide based on layouts, but not
functorially. And finally, we highlighted that the proof of
the Totality \theoremref{totality} makes use of the fact that
initialisation data can be promoted along world extension. We will now
expose this structure semantically.

\subsection{Worlds}

The category $\worlds$ of \emph{worlds} has as objects the heap
layouts of \secref{full ground storage}, i.e., partial functions $\w :
\locs \pfinto \cellnames$ with finite support $\ord\w \subset \locs$. A
morphism $\h : \w \to \w'$ is an injection $\h : \ord\w \inject
\ord\w'$ such that $(\ell : \Cll) \in \w$ implies $(\h(\ell) : \Cll)
\in \w'$. For brevity, we refer to heap layouts as worlds and to
$\worlds$-morphisms as (world) injections.

We define several layout-manipulation operations on $\worlds$. While
we work concretely, these operations can be axiomatised by universal
properties using Simpson's \emph{independence
  structures}~\cite{simpson:independence-structures}, and we will use his vocabulary as much as possible.

Let $\# : \naturals \xto{\isomorphic}\locs$ be an enumeration of the
set of locations. Given a world $\w$, define $\numof\w$ to be the
smallest index beyond all the locations in $\w$, i.e.~${\min\set*{n \in
    \naturals \suchthat* \forall i \geq n. \lnum i \notin \w}}$.
Given two worlds $\w_1, \w_2 \in \worlds$, we can embed their supports
into the following subset of $\locs$:
\[
\ord{\w_1\+\w_2} \definedby
\ord\w_1\union\set*{\lnum{(\numof{\w_1} + n)}\suchthat* \lnum n \in \w_2}
\]
by setting $\iinj_1(\ell) \definedby \ell$ and $\iinj_2(\lnum n)
\definedby \lnum(\numof {\w_1} + n)$. We then have that for every $\ell \in
\ord{\w_1\+\w_2}$ there is exactly one $i \in \set{1,2}$ and ${\ell_i
  \in \ord\w_i}$ such that $\ell = \iinj_i\ell_i$. We define the
\emph{independent coproduct} $\w_1\+\w_2$ whose support is given by
$\ord{\w_1\+\w_2}$ by setting \( (\w_1\+\w_2)(\iinj_i\ell) \definedby
\w_i(\ell) \). Then $\iinj_i : \w_i \to \w_1\+\w_2$ are world
injections which we call the \emph{independent coprojections}. Moreover, the
construction $\+$ extends to a functor $\+ : \worlds\times\worlds
\to\worlds$ and each coprojection is a natural transformation. The
independent coproduct is \emph{not} a coproduct in $\worlds$, for
example, there is no codiagonal injection $\w\+\w \to \w$ for $\w =
\set{\ell : \Cll}$. Independent coproducts are the semantic
counterparts to extending a world with fresh locations.

Given an injection $\h : \w_1 \to \w_2$, its \emph{complement} is the
injection $\comp\h : \w_2\o-\h \to \w_2$ whose domain
${\ord{\w_2\o-\h} := \ord\w_2 \setminus \Image\h}$ are all the
locations in $\w_2$ that $\h$ misses, and the action of $\comp\h$ is
given by that of $\w_2$. There are canonical isomorphisms
$\w_1\+(\w_2\o-\h) \isomorphic \w_2$ and $(\w_1\+\w_2)\o-\iinj_i
\isomorphic \w_{3-i}$. We use complements to define initialisation
data below.

Given two injections $\h_i : \w \to \w_i$, $i=1,2$, we define their
\emph{local independent coproduct} by \[
\h_1\o+ \h_2 := \w \+ (\w_1 \o- \h_1) \+ (\w_2 \o- \h_2)
\]
We have morphisms $\w_1 \xto{\h_{1}\mis'\h_{2}} \h_1\o+ \h_2 \xfrom{\h_{2}\mis\h_{1}} \w_2 $
such that:
\cdiagram{local-coproducts-01}

We define the functor category $\vals := [\worlds, \Set]$ in which we
will interpret the types of the $\lamref$-calculus. We interpret the
\emph{full ground types}, defining $\sem\placeholder : \grounds \to
\vals$ by:
\begin{mathpar}
  {\sem{\s0} \definedby \initial

  \sem{\gty_1\s+\gty_2} \definedby \sem{\gty_1}+\sem{\gty_2}}

  {\sem{\s1} \definedby \terminal\,

  \sem{\gty_1\s*\gty_2} \definedby \sem{\gty_1}\times\sem{\gty_2}}

  {\sem{\sref \Cll}w \definedby \set{\ell \in \w \suchthat \w(\ell) = \Cll}

  (\sem{\sref \Cll}\h)\ell \definedby \h(\ell)}
\end{mathpar}
We can interpret references more compactly by noting that
$\sem{\sref\Cll} \isomorphic \worlds(\set{\ell : \Cll}, \placeholder)$.

\subsection{Initialisations}
The account so far has been standard for possible-world semantics of
local state. We now turn to defining the semantic counterpart to initialisation data.

Define the mixed-variance functor $\prostores : \opposite\worlds\times\worlds \to \Set$:
\[
\prostores(\w^-, \w^+) \definedby \prod_{(\ell : \Cll) \in \w^-} \sem{\typeof \Cll}\w^+
\]
Its contravariant action is given by projection, and its covariant
action is given component-wise by the actions of $\sem{\typeof\Cll}$.
Elements of $\prostores(\w^-, \w^+)$ are
\emph{heaplets}~\cite{ohearn:scalable-specification-and-reasoning}
whose layout is given by $\w^-$, and whose values assume the layout
$\w^+$~(\subfigref{heaplet def}). As in separation logic, heaplets are
a composable abstraction facilitating local reasoning about the
heap. This functor preserves the independent coproducts in the sense
that $\prostores(\w_1\+\w_2, \w)$ and
\( {\prostores(\iinj_i, \w) : \prostores(\w_1\+\w_2, \w) \to
\prostores(\w_i, \w)} \) form the product of $\prostores(\w_1, \w)$ and
$\prostores(\w_2, \w)$, and that $\prostores(\emptyset, \w)$ is the
singleton. Consequently, we have canonical
isomorphisms \( \storesuni : \terminal \xto{\isomorphic}
\prostores(\emptyset, \w) \) and, depicted in \subfigref{heaplet
  concat},
\( \storesmon : \prostores(\w_1, \w)\times \prostores(\w_2, \w)
\xto{\isomorphic} \prostores(\w_1\+\w_2, \w) \). \subfigref{heaplet
  defrag} depicts the contravariant action of $\prostores$ on the
canonical isomorphism $\w_2 \isomorphic \w_1\+(\w_2\o-\h)$.

The category $\init$ of \emph{initialisations} has worlds as objects,
and as homsets $\init(\w_1, \w_2) \definedby \sum_{\h : \w_1 \to \w_2}
\prostores(\w_2\o-\h, \w_2)$ whose elements we call
\emph{initialisations}. Explicitly, an initialisation ${\ih : \w_1 \to
  \w_2}$ is a pair $\pair{\p\ih}{\store_{\ih}}$ consisting of an
injection $\p\ih : \w_1 \to \w_2$ and a heaplet $\store_{\ih}$
containing the initialisation data required to transition from heap
layout $\w_1$ to $\w_2$~(\subfigref{init data def}). This heaplet may contain cyclic dependencies on
the newly added locations, or on locations already present in
$\w_1$. Identities $\id[\w]^{\init}$ in $\init$ are given by identities in
$\worlds$, and formally as $\pair{\id[\w]^{\worlds}}\storesuni$, as no
initialisation data is required. The composition of two
initialisations is given by composing their underlying injections, and
appending their initialisation data, suitably promoted to the later
world~(\subfigref{init data composition}).

The collection of \emph{(semantic) heaps} now becomes a representable
functor $\stores : \init \to \Set$, given at world $\w$ by setting
\[
\stores\w \definedby \prostores(\w, \w) \isomorphic \init(\emptyset, \w)
\]
The latter bijection follows from the canonical isomorphism
$\w\o-\id[\w] \isomorphic \emptyset$ in $\worlds$. The functorial
action of $\init(\emptyset, \placeholder)$ then equips $\stores$ with
a functorial action over initialisations: given a heap $\store \in
\stores\w_1$ and an initialisation $\ih : \w_1 \to \w_2$, promote
$\store$ to a heaplet in $\stores(\w_1, \w_2)$, and append the
initialisation data to create a heap in $\stores\w_2$. Given $(\ell :
\Cll) \in \w$, we use projection to define a look-up operation given
for any $\store \in \stores\w$ by setting $\store(\ell) \definedby
\projection_{\ell}\store$, and an update operation, given for any
$\store \in \stores\w$ and $x \in \sem{\typeof \Cll}\w$ by setting
\[
\store[\ell \mapsto x](\ell') \definedby \begin{cases}
  x              & \ell' = \ell \\
  \store(\ell')  & \text{otherwise}
\end{cases}
\]

\begin{wrapfigure}{r}{.22\textwidth}
\vspace{-.3cm}
\begin{tikzpicture}[x=.25in,y=0.3cm, >=stealth]

  \foreach \i in {0}
  {
    \DashedHeapCell{(UL\i)}{(ULr\i)}{(0,-\i)}
    \patternA*{(UL\i)}{(ULr\i)}
  }

  \foreach \i in {1,2}
  {
    \DashedCellDownwards{(UL\i)}{(ULr\i)}{(0,-\i)}
    \patternA*{(UL\i)}{(ULr\i)}
  }

     \node[above left] at (UL0) {$\w_1$};

  \foreach \i in {0,1}
  {
    \DashedCellUpwards{(UL\i)}{(ULr\i)}{(3.5,-\i)}
    \patternA*{(UL\i)}{(ULr\i)}
  }

  \foreach \i in {2}
  {
    \HeapCell{(UR)}{(URr)}{(3.5, -\i)}
    \patternB{(UR)}{(URr)}{(3.5, -\i)}
  }

  \foreach \i in {2}
  {
    \DashedCellDownwards{(UL\i)}{(ULr\i)}{(3.5,-1-\i)}
    \patternA*{(UL\i)}{(ULr\i)}
  }

  \draw[*->] let \p1 = ($(UR)!0.5!(URr)+(+2pt, 0)$),
                 \p2 = ($(UL1)!0.5!(ULr1)- (.5, 0) - (2pt, 1pt)$)
             in (\x1, \y1) -- (\x1 - 15pt - 1*5pt, \y1)
                           -- (\x1 - 15pt - 1*5pt, \y2)
                           -- (\x2, \y2);

     \node[above left] at (UL0) {$\store_{\ih}$};

  \foreach \i in {0}
  {
    \DashedCellUpwards{(UL\i)}{(ULr\i)}{(0,-5-\i)}
    \patternA*{(UL\i)}{(ULr\i)}
  }

  \foreach \i in {1}
  {
    \DashedHeapCell{(foob)}{(foobr)}{(0, -5-\i)}
  }

  \foreach \i in {1,2}
  {
    \DashedCellDownwards{(UL\i)}{(ULr\i)}{(0,-6-\i)}
    \patternA*{(UL\i)}{(ULr\i)}
  }

  \node[above left] at (UL0) {$\w'$};

  \foreach \i in {0}
  {
    \DashedCellUpwards{(UL\i)}{(ULr\i)}{(3.5,-5-\i)}
    \patternA*{(UL\i)}{(ULr\i)}
  }

  \foreach \i in {1,3}
  {
    \DashedHeapCell{(foob)}{(foobr)}{(3.5, -5-\i)}
  }

  \foreach \i in {1,2}
  {
    \DashedCellDownwards{(UL\i)}{(ULr\i)}{(3.5,-5-\i)}
    \patternA*{(UL\i)}{(ULr\i)}
  }

  \foreach \i in {4}
  {
    \HeapCell{(UR)}{(URr)}{(3.5, -5-\i)}
    \patternB{(UR)}{(URr)}{(3.5, -5-\i)}
  }

  \draw[*->] let \p1 = ($(UR)!0.5!(URr)+(+2pt, 0)$),
                 \p2 = ($(UL1)!0.5!(ULr1)- (.5, 0) - (2pt, 1pt)$)
             in (\x1, \y1) -- (\x1 - 15pt - 1*5pt, \y1)
                           -- (\x1 - 15pt - 1*5pt, \y2)
                           -- (\x2, \y2);

   \node[above left] at (UL0) {$\store_{\h\mis\ih}$};
\end{tikzpicture}
\caption{Promoting init.~data}\figlabel{unsoundness}
\vspace{-.25cm}
\end{wrapfigure}
Finally, given any injection $\h : \w_1 \to \w'$ and initialisation
$\ih : \w_1 \to \w_2$, the injection $\h\mis'\p\ih : \w' \to
\h\o+[\w_1]\p\ih$ in fact has an initialisation structure $\h\mis\ih :
\w' \to \h\o+[\w_1]\p\ih$, where the initialisation data
$\store_{\h\mis\ih}$ is given using the isomorphism
\[
(h\o+[\w_1]\p\ih)\o-\h\mis\p\ih \isomorphic \w_2\o-\p\ih
\]
and promotion along $\p\ih\mis\h$. We denote $\p\ih\mis\h$ by
$\ih\mis\h$. This process is the semantic counterpart for the
promotion of initialisation data we use in the proof of the Totality
\theoremref{totality}.


\section{The monad}\seclabel{monad}
Consider the functor category $\pconfs := [\init, \Set]$, which
contains the heaps functor as an object.  As we have a forgetful
functor $\p : \init \to \worlds$ projecting out the underlying
injection, we obtain a functor $\p^* : \vals \to \pconfs$ given by
precomposition. In the following, consider the cartesian closed
structure of $\vals$ as a symmetric monoidal closed structure.

We equip $\pconfs$ with a bi-closed $\vals$-actegory structure:
\begin{mathpar}
    X\.A := \p^*X\times A := (X\compose\p)\times A

    A\acthom B := \p_*\parent{B^A}

    X\powers A := A^{\p^*X}

    \compact_{X,Y,A} : \pair{\pair xy}a \mapsto \pair{x}{\pair ya}

    \trivact_{A} : \pair\star{a} \mapsto a
\end{mathpar}
This structure can be alternatively described as transporting the
self-enrichment of $\pconfs$ via the cartesian closed structure along
the geometric morphism $\pair{\p^*}{\p_*}$ from $\pconfs$ to $\vals$.

We can give an explicit end formula for the enrichment:
\begin{mathpar}
  (A\acthom B)\w := \cocoend{\w \to \w' \in \w\comma\p}(B\w')^{A\w'}

  \projection_{\h' : \w_2 \to \w_2'} (A\acthom B)(\vto{\w_1}\h{\w_2})(\alpha) :=
  \projection_{\h' \compose \h}\alpha

  \projection_{\h : \w \to \w'} (\curry_{X, A, B}^{\acthom} f(x))(a) = f_{\w'}(X\h x, a)

  (\eval^{\acthom}_{X, A})_{\w} (\alpha, a)) = \projection_{\id[\w]}\alpha \ a
\end{mathpar}

We can now give an explicit description of the full ground storage
monad $T : \vals \to \vals$. The action on worlds is
\[
  (TX)\w \definedby
      \cocoend{\w \to \w' \in  \w\comma\p}
         \parent{\coend*{\w' \to \w'' \in \w\comma\p }
           \quad (X\compose\p)\w''\times\stores\w''}^{\stores\w'}
\]
This definition is subtle. First, the argument of the coend is
covariant in $\w'\to\w'' \in \w\comma\p$, and so this coend is an
ordinary colimit. We keep the coend notation for its more
convenient presentation.  The second subtlety is that, while the inner
coend is contravariant in $\w'$, the action with respect to which we
define the outer end is \emph{different}, and is in fact
\emph{co}variant in the object $\w\to\w'$ of the comma category
$\w\comma\p$. To describe it explicitly, take any morphism
$\ih : \pair {\w'_1}{\h'_1} \to \pair{\w'_2}{\h'_2}$ in the comma
category, i.e., an initialisation $\ih : \w'_1 \to \w'_2$ such that
$\p\ih\compose \h'_1 = \h'_2$. Consider a generic element in the coend
\( \coend{\w'_1 \to \w'' \in \w\comma\p }
(X\compose\p)\w''\times\stores\w'' \) namely some $q_{\h}(x, \store)$,
for some $\h : \w'_1 \to \w''$, $x \in X\w''$ and
$\store \in \stores\w''$. We promote the initialisation $\ih$ to an
initialisation $\h\mis\ih : \w'' \to \h\o+[\w'_1]\p\ih$, and map the
generic element as follows:
\[
q_{\h}(x, \store) \mapsto q_{\p\ih\mis\h : \w_2' \to \h\o+[\w'_1]\p\ih}(X(\p(\h\mis\ih))x, \stores(\h\mis\ih)\store)
\]
This subtlety is the main conceptual reason for the decomposition of
this monad we present in the next section.  We do indeed use the
contravariant action of the coend, implicitly below, and explicitly in
the next section, to define the hiding/encapsulation operation.  This
subtlety also appears in the (ordinary) ground storage
monad~\cite{plotkin-power:notions-of-computation-determine-monads}
when defining the functorial action of $TX$. The end gives the
functorial action in the full ground setting:
\[
(\projection_{\h_2 : \w_2 \to \w'_2}(TX(\vto{\w_1}{\h}{\w_2})\alpha))(\store_2) = \projection_{\h_2\compose \h}(\alpha)(\store_2)
\]
The monadic unit is given by \(
(\projection_{\h : \w \to \w'}\compose \mreturn^T_{\w} x)\store \definedby q_{\id[\w']}(X\h x, \store)
\).
Given any morphism $f : X \to TY$ in $\vals$ and $\alpha \in TX$, define
\(
(\projection_{\h : \w\to\w'}(\alpha\bind f))(\store') = q_{\h'' \compose \h'}(y, \store''')
\)
where
\begin{align*}
(\projection_{\h}\alpha)\store' &= q_{\h' : \w' \to \w''}(x,
\store'')
\\
(\projection_{\id[\w'']}\compose f_{w''}(x))(\store'')
&= q_{\h'' : \w'' \to \w'''}(y, \store''')
\end{align*}
Define the strength for any $x \in X\w$ and $\alpha \in TX\w$:
\[
(\projection_{\h : \w\to\w'}\compose\strength^T_{\w}(x, \alpha))\store' = q_{\h'}(\pair{X(\h'\compose\h)x}y, \store'')
\]
where $(\projection_{\h}\alpha)\store' = q_{\h'' : \w' \to \w''}(y,
\store'')$.  Finally, from the other definitions we calculate the
functorial action of $T$ on any morphism $f : X \to Y$:
\(
(\projection_{\h : \w \to \w}(Tf\alpha))\store' = q_{\h'}(f_{\w''}(x), \store'')
\)
where $(\projection_{\h}\alpha)\store' = q_{\h' : \w' \to \w''}(x, \store'')$.

On this monad we define the state manipulation operations by setting,
for every $\h : \w \to \w'$, two $\pconfs$-morphisms:
\[
\begin{array}{*4{@{}l}}
  \mget_{\Cll} &{}:{}& \multicolumn2{@{}l}{\sem{\sref \Cll} \to T \sem {\typeof\Cll}}\\
  &&(\projection_{\h}\compose\mget_{\Cll}(\ell))(\store_1)
  &=
  q_{\id[\w']}(\store(\h(\ell)), \store)
  \\
  \mset_{\Cll} &{}:{}& \multicolumn2{@{}l}{\sem{\sref \Cll} \times \sem {\typeof \Cll} \to T\terminal
  }\\
  &&(\projection_{\h}\compose\mset_{\Cll}(\ell, a))(\store_1)
  &=
  q_{\id[\w']}(\star, \store[\h(\ell) \mapsto \sem{\typeof\Cll}\h a])
\end{array}
\]
To define the allocation operation, first define, for every $\w_0$ in
$\worlds$ the functor $\mshift_{w_0} : \vals \to \vals$ that evaluates
at a later world, namely $\mshift_{\w_0}X \definedby X(\placeholder \+
\w_0)$.  Using the isomorphism $\phi : (\w\+\w_0)\o-\iinj_1\isomorphic \w_0$,
we can then define the $\vals$-morphism that constructs an initialisation from
given initialisation data:
\[
\begin{array}{*4{@{}l}}
  \minit_{\w_0, \w} &{}:{}& \multicolumn2{@{}l}{
    \prod_{(\ell : \Cll) \in \w_0} \mshift_{\w_0}\sem{\typeof \Cll}\w \to \init(\w, \w\+\w_0)
  }\\
  &&\seq[\ell \in \ord\w_0]{a_\ell} \mapsto \pair{\iinj_1\compose \phi}{\seq[\ell]{a_{\phi\ell}}}
\end{array}
\]
and define:
\[
\begin{array}{*4{@{}l}}
  \mnew_{\w_0} &:{}& \multicolumn2{@{}l}{
    \prod_{(\ell : \Cll) \in \w_0} \mshift_{\w_0}\!\sem{\typeof\Cll}
    \to
    T\prod_{(\ell : \Cll)\in\w_0}\sem{\sref \Cll}
  }\\
  &&
  (\projection_{\h}\compose\mnew_{\w_0}\seq{a_{\ell}})\store_1 &=
q_{\ih\mis\h}(\seq[\ell \in \w_0]{\ih\mis\h(\ell)},
\stores(\h\mis\ih)\store_1)
\end{array}
\]
where $\ih \definedby \minit\seq{a_{\ell}}$.


\section{Hiding and masking}\seclabel{hiding}
\begin{figure*}
\hfil%
\subfloat[{$\smash{PA\w_2 \xto{\mhide_\h } PA\w_1}$}]{%
\begin{tikzpicture}[x=.25in,y=0.3cm, >=stealth]

  \def\offset{0}

  \foreach \i in {0, 1, 2}
  {
    \HeapCell{(Cell\i)}{(Cellr\i)}{(\offset, -\i)}
  }

  \patternA{(Cell2)}{(Cellr2)}

  \foreach \i in {3,4}
  {
    \HeapCell{(Cell\i)}{(Cellr\i)}{(\offset, -\i-.5)}
  }

  \foreach \i/\j in {1/2, 2/3, 3/4}
  {
  \draw[*->] let \p1 = ($(Cell\i)!0.5!(Cellr\i)+(-1pt, 0)$),
                 \p2 = ($(Cell\j)!0.5!(Cellr\j)+ (.5, 0) + (2pt, 2pt)$)
             in (\x1, \y1) -- (\x1 + 15pt + .25*5pt, \y1)
                           -- (\x1 + 15pt + .25*5pt, \y2)
                           -- (\x2, \y2);
  };

  \draw[dashed] let \p1 = ($(Cell4)+(-4pt, -.25)$),
                    \p2 = ($(Cellr3)+(10pt,  .25)$)
                in  (\x1, \y1) -- (\x2, \y1)
                               -- (\x2, \y2)
                               -- (\x1, \y2)
                               -- cycle;

  \draw[|->] let \p1 = ($(Cell0)!0.5!(Cellr0)+(-1pt, 0)$),
                 \p2 = ($(Cell2)!0.5!(Cellr2)+ (.5, 0) + (2pt, 2pt)$) in
                 (\x2+10, \y2-5pt) -- (\x2+20, \y2-5pt);


  \def\offset{2.6}

  \foreach \i in {0, 1}
  {
    \HeapCell{(Cell\i)}{(Cellr\i)}{(\offset, -\i)}
  }

  \foreach \i in {2, 3,4}
  {
    \HeapCell{(Cell\i)}{(Cellr\i)}{(\offset, -\i-.5)}
  }

  \patternA{(Cell2)}{(Cellr2)}

  \foreach \i/\j in {1/2, 2/3, 3/4}
  {
  \draw[*->] let \p1 = ($(Cell\i)!0.5!(Cellr\i)+(-1pt, 0)$),
                 \p2 = ($(Cell\j)!0.5!(Cellr\j)+ (.5, 0) + (2pt, 2pt)$)
             in (\x1, \y1) -- (\x1 + 15pt + .25*5pt, \y1)
                           -- (\x1 + 15pt + .25*5pt, \y2)
                           -- (\x2, \y2);
  };

  \draw[dashed] let \p1 = ($(Cell4)+(-4pt, -.25)$),
                    \p2 = ($(Cellr2)+(10pt,  .25)$)
                in  (\x1, \y1) -- (\x2, \y1)
                               -- (\x2, \y2)
                               -- (\x1, \y2)
                               -- cycle;

\end{tikzpicture}
\subfiglabel{hiding hide}}%
\hfil%
\subfloat[Return]{%
\begin{tikzpicture}[x=.25in,y=0.3cm, >=stealth]

  \def\offset{0}

  \foreach \i in {0, 1, 2}
  {
    \HeapCell{(Cell\i)}{(Cellr\i)}{(\offset, -\i)}
  }

  \foreach \i/\j in {0/1}
  {
  \draw[*->] let \p1 = ($(Cell\i)!0.5!(Cellr\i)+(-1pt, 0)$),
                 \p2 = ($(Cell\j)!0.5!(Cellr\j)+ (.5, 0) + (2pt, 2pt)$)
             in (\x1, \y1) -- (\x1 + 15pt + .25*5pt, \y1)
                           -- (\x1 + 15pt + .25*5pt, \y2)
                           -- (\x2, \y2);
  };

  \draw[|->] let \p1 = ($(Cell0)!0.5!(Cellr0)+(-1pt, 0)$),
                 \p2 = ($(Cell1)!0.5!(Cellr1)+ (.5, 0) + (2pt, 2pt)$) in
                 (\x2+5, \y2-5pt) -- (\x2+15, \y2-5pt);


  \def\offset{2}

  \foreach \i in {0, 1, 2}
  {
    \HeapCell{(Cell\i)}{(Cellr\i)}{(\offset, -\i)}
  }

  \foreach \i/\j in {0/1}
  {
  \draw[*->] let \p1 = ($(Cell\i)!0.5!(Cellr\i)+(-1pt, 0)$),
                 \p2 = ($(Cell\j)!0.5!(Cellr\j)+ (.5, 0) + (2pt, 2pt)$)
             in (\x1, \y1) -- (\x1 + 15pt + .25*5pt, \y1)
                           -- (\x1 + 15pt + .25*5pt, \y2)
                           -- (\x2, \y2);
  };

  \draw[dashed] let \p1 = ($(Cell2)+(-4pt, -1.5)$),
                    \p2 = ($(Cellr2)+(+4pt,-2+.5)$)
                in  (\x1, \y1) -- (\x2, \y1)
                               -- (\x2, \y2)
                               -- (\x1, \y2)
                               -- cycle;

\end{tikzpicture}
\subfiglabel{hiding return}}%
\hfil%
\subfloat[\hbox{Bind: $\smash{\stores\w'\xto{g_{\w'}}P\stores\w'}$
(left), $\smash{\stores\w\xto{\protect\bind g} \stores\w}$ (right)}]{%
\begin{tikzpicture}[x=.25in,y=0.3cm, >=stealth]

  \def\offset{0}

  \foreach \i in {0, 1, 2}
  {
    \HeapCell{(Cell\i)}{(Cellr\i)}{(\offset, -\i)}
  }

  \patternA{(Cell2)}{(Cellr2)}

  \foreach \i/\j in {1/2}
  {
  \draw[*->] let \p1 = ($(Cell\i)!0.5!(Cellr\i)+(-1pt, 0)$),
                 \p2 = ($(Cell\j)!0.5!(Cellr\j)+ (.5, 0) + (2pt, 2pt)$)
             in (\x1, \y1) -- (\x1 + 15pt + .25*5pt, \y1)
                           -- (\x1 + 15pt + .25*5pt, \y2)
                           -- (\x2, \y2);
  };

  \draw[|->] let \p1 = ($(Cell0)!0.5!(Cellr0)+(-1pt, 0)$),
                 \p2 = ($(Cell1)!0.5!(Cellr1)+ (.5, 0) + (2pt, 2pt)$) in
                 (\x2+7.5, \y2-5pt) -- (\x2+15, \y2-5pt);


  \def\offset{2}

  \foreach \i in {0, 1,2}
  {
    \HeapCell{(Cell\i)}{(Cellr\i)}{(\offset, -\i)}
  }

  \foreach \i in {3,4}
  {
    \HeapCell{(Cell\i)}{(Cellr\i)}{(\offset, -\i-.5)}
    \patternB{(Cell\i)}{(Cellr\i)}
  }

  \patternA{(Cell2)}{(Cellr2)}

  \foreach \i/\j/\k/\r in {1/3/-2/2, 3/2/0/1}
  {
  \draw[*->] let \p1 = ($(Cell\i)!0.5!(Cellr\i)+(-1pt, 0)$),
                 \p2 = ($(Cell\j)!0.5!(Cellr\j)+ (.5, 0) + (2pt, \k pt)$)
             in (\x1, \y1) -- (\x1 + 15pt + \r*5pt, \y1)
                           -- (\x1 + 15pt + \r*5pt, \y2)
                           -- (\x2, \y2);
  };

  \draw[dashed] let \p1 = ($(Cell4)+(-4pt, -.25)$),
                    \p2 = ($(Cellr3)+(+6pt,  .25)$)
                in  (\x1, \y1) -- (\x2, \y1)
                               -- (\x2, \y2)
                               -- (\x1, \y2)
                               -- cycle;

\end{tikzpicture}
\qquad
\begin{tikzpicture}[x=.25in,y=0.3cm, >=stealth]

  \def\offset{0}
  \foreach \i in {0, 1}
  {
    \HeapCell{(Cell\i)}{(Cellr\i)}{(\offset, -\i)}
  }

  \HeapCell{(Cell2)}{(Cellr2)}{(\offset, -2 -.5)}

  \patternA{(Cell2)}{(Cellr2)}

  \foreach \i/\j in {1/2}
  {
  \draw[*->] let \p1 = ($(Cell\i)!0.5!(Cellr\i)+(-1pt, 0)$),
                 \p2 = ($(Cell\j)!0.5!(Cellr\j)+ (.5, 0) + (2pt, 2pt)$)
             in (\x1, \y1) -- (\x1 + 15pt + .5*5pt, \y1)
                           -- (\x1 + 15pt + .5*5pt, \y2)
                           -- (\x2, \y2);
  };

  \draw[dashed] let \p1 = ($(Cell2)+(-4pt, -.25)$),
                    \p2 = ($(Cellr2)+(+6pt,  .25)$)
                in  (\x1, \y1) -- (\x2, \y1)
                               -- (\x2, \y2)
                               -- (\x1, \y2)
                               -- cycle;

  \draw[|->] let \p1 = ($(Cell0)!0.5!(Cellr0)+(-1pt, 0)$),
                 \p2 = ($(Cell1)!0.5!(Cellr1)+ (.5, 0) + (2pt, 2pt)$) in
                 (\x2+10, \y2-5pt) -- (\x2+20, \y2-5pt);

  \def\offset{2.5}

  \foreach \i in {0, 1}
  {
    \HeapCell{(Cell\i)}{(Cellr\i)}{(\offset, -\i)}
  }

  \foreach \i in {2,3,4}
  {
    \HeapCell{(Cell\i)}{(Cellr\i)}{(\offset, -\i-.5)}
  }

  \foreach \i in {3,4}
  {
    \patternB{(Cell\i)}{(Cellr\i)}
  }

  \patternA{(Cell2)}{(Cellr2)}

  \foreach \i/\j/\k/\r in {1/3/-2/2, 3/2/0/1}
  {
  \draw[*->] let \p1 = ($(Cell\i)!0.5!(Cellr\i)+(-1pt, 0)$),
                 \p2 = ($(Cell\j)!0.5!(Cellr\j)+ (.5, 0) + (2pt, \k pt)$)
             in (\x1, \y1) -- (\x1 + 15pt + \r*5pt, \y1)
                           -- (\x1 + 15pt + \r*5pt, \y2)
                           -- (\x2, \y2);
  };

  \draw[dashed] let \p1 = ($(Cell4)+(-4pt, -.25)$),
                    \p2 = ($(Cellr2)+(+6pt,  .25)$)
                in  (\x1, \y1) -- (\x2, \y1)
                               -- (\x2, \y2)
                               -- (\x1, \y2)
                               -- cycle;

\end{tikzpicture}
\subfiglabel{hiding bind}}%
\hfil%
\subfloat[Functorial action (derived)]{
\begin{tikzpicture}[x=.25in,y=0.3cm,>=stealth]
  \def\offset{-2.2}
  \def\Strut{\vphantom{\begin{aligned}~\\~\\~\end{aligned}}}
  \foreach \i in {0,1}
  {
    \DashedCellUpwards{(UL\i)}{(ULr\i)}{(\offset,-\i)}
    \patternA*{(UL\i)}{(ULr\i)}
  }

  \foreach \i in {2}
  {
    \HeapCell{(UR)}{(URr)}{(\offset, -\i)}
    \patternB{(UR)}{(URr)}{(\offset, -\i)}
  }

  \foreach \i in {2}
  {
    \DashedCellDownwards{(UL\i)}{(ULr\i)}{(\offset,-1-\i)}
    \patternA*{(UL\i)}{(ULr\i)}
  }

  \draw[*->] let \p1 = ($(UR)!0.5!(URr)+(+2pt, 0)$),
                 \p2 = ($(UL1)!0.5!(ULr1)- (.5, 0) - (2pt, 1pt)$)
             in (\x1, \y1) -- (\x1 - 15pt - 1*5pt, \y1)
                           -- (\x1 - 15pt - 1*5pt, \y2)
                           -- (\x2, \y2);

  \node[above left] at (UL0) {$\store_{\ih}$};
  \node[left] at ($(UL1) + (-.5, 0)$) {$P\left(\Strut\right.$};
  \node[left] at ($(UL1) + (2.1, 0)$) {$\left)\Strut\!\!\right($};
\node[left] at ($(UL1) + (5.3, 0)$) {$\left.\Strut\right)=$};

  \foreach \i in {0}
  {
    \HeapCell{(UL\i)}{(ULr\i)}{(0,-\i+.5)}
    \patternA{(UL\i)}{(ULr\i)}
  }

  \foreach \i in {1}
  {
    \HeapCell{(UL\i)}{(ULr\i)}{(0,-\i)}
  }

  \foreach \i in {2,3}
  {
    \HeapCell{(UL\i)}{(ULr\i)}{(0,-\i-.5)}
    \patternA{(UL\i)}{(ULr\i)}
  }

  \draw[*->] let \p1 = ($(UL1)!0.5!(ULr1)+(-1pt, 0)$),
                 \p2 = ($(UL2)!0.5!(ULr2)+ (.5, 0) + (2pt, 1pt)$)
             in (\x1, \y1) -- (\x1 + 15pt + 1*5pt, \y1)
                           -- (\x1 + 15pt + 1*5pt, \y2)
                           -- (\x2, \y2);

  \draw[*->] let \p1 = ($(UL3)!0.5!(ULr3)+(-1pt, 0)$),
                 \p2 = ($(UL1)!0.5!(ULr1)+ (.5, 0) + (2pt, 2pt)$)
             in (\x1, \y1) -- (\x1 + 15pt + 2*5pt, \y1)
                           -- (\x1 + 15pt + 2*5pt, \y2)
                           -- (\x2, \y2);

  \draw[dashed] let \p1 = ($(UL1)+(-8pt, -.25)$),
                    \p2 = ($(ULr1)+(8pt,  .25)$)
                in  (\x1, \y1) -- (\x2, \y1)
                               -- (\x2, \y2)
                               -- (\x1, \y2)
                               -- cycle;


  \foreach \i in {3.1}
  {
    \HeapCell{(foob)}{(foobr)}{(3, -\i)}
  }

  \foreach \i in {0,1,2}
  {
    \HeapCell{(UL\i)}{(ULr\i)}{(3,-\i+.5)}
    \patternA{(UL\i)}{(ULr\i)}
  }

  \foreach \i in {4}
  {
    \HeapCell{(UR)}{(URr)}{(3, -\i-.5)}
    \patternB{(UR)}{(URr)}{(3, -\i-.5)}
  }

  \draw[*->] let \p1 = ($(UR)!0.5!(URr)+(-1pt, 0)$),
                 \p2 = ($(UL1)!0.5!(ULr1)+ (.5, 0) + (2pt, +1.5pt)$)
             in (\x1, \y1) -- (\x1 + 15pt + 2*5pt, \y1)
                           -- (\x1 + 15pt + 2*5pt, \y2)
                           -- (\x2, \y2);


  \draw[*->] let \p1 = ($(foob)!0.5!(foobr)+(-1pt, 0)$),
                 \p2 = ($(UL1)!0.5!(ULr1)+ (.5, 0) + (2pt, -1.5pt)$)
             in (\x1, \y1) -- (\x1 + 15pt + 1*5pt, \y1)
                           -- (\x1 + 15pt + 1*5pt, \y2)
                           -- (\x2, \y2);

  \draw[*->] let \p1 = ($(UL2)!0.5!(ULr2)+(-1pt, 0)$),
                 \p2 = ($(foob)!0.5!(foobr)+ (.5, 0) + (2pt, 2pt)$)
             in (\x1, \y1) -- (\x1 + 15pt + .25*5pt, \y1)
                           -- (\x1 + 15pt + .25*5pt, \y2)
                           -- (\x2, \y2);

  \draw[dashed] let \p1 = ($(foob)+(-8pt, -.25)$),
                    \p2 = ($(foobr)+(8pt,  .25)$)
                in  (\x1, \y1) -- (\x2, \y1)
                               -- (\x2, \y2)
                               -- (\x1, \y2)
                               -- cycle;

\end{tikzpicture}
\subfiglabel{hiding functorial}}%
\hfil%
\caption{The hiding monad $P$}
\figlabel{hiding monad}
\end{figure*}
We now analyse the functorial action of the inner coend in $T$'s
definition, which is given by a $\moncat\vals$-strong monad
$\structure P$ on $\pconfs$.

\subsection{The hiding monad}
Consider any $A \in \pconfs$, and define
for every $\w$:
\[
PA\w \definedby \coend{\w \to \w' \in \w\comma\p}A
\]
Given an extension $\h : \w \to \w'$, we think of locations in
$\w'\o-\h$ as \emph{private} locations, and of locations in $\w$ as
\emph{public} locations.
\begin{example}
  On the left we depict two representatives for a value in
  $P\stores\set{\ell:\data}$. The left representative has no private
  locations, whereas the right representative has the two private
  locations $\set{\ell_0 : \linkedlist, \ell_1 : \cell}$. As we can
  initialise the right representative from the left, the two
  representatives are equivalent.

  
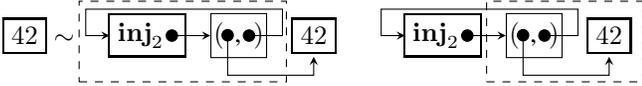
\begin{center}
\begin{tikzpicture}[list/.style={rectangle split, rectangle split parts=2,
  draw, rectangle split horizontal, rectangle split draw splits=false}, >=stealth, start chain,
  node distance=.20cm]

\node[on chain]
{\hspace{-.4cm}$\framebox{$42$}\sim$};
\node[on chain] (C) {\framebox{$\sinj[]2\hphantom{\bullet}$}};
\node[list,on chain] (A) {\,$\mathllap{(}$\hspace{10pt}\nodepart{two}\hspace{-3pt},\hspace{4.5pt}$\mathrlap{)}$\,};
\node[on chain] (B) {\framebox{$42$}};

\draw[*->] let \p1 = (C.center), \p2 = (A) in (\x1 + 7, \y1) -- (A);
\draw[*->] let \p1 = (A.two), \p2 = (A.center), \p3 = (C.west) in
(\x1,\y2) -- (\x1 + 15, \y2) -- (\x1 + 15, \y2 + 10) -- (\x3 - 5, \y2 + 10)
          -- (\x3 - 5, \y2) -- (\x3+4, \y3);
\draw[*->] let \p1 = (A.one), \p2 = (B) in (\x1+3,\y1 + 5) -- (\x1 + 3, \y1 - 12.5) -- (\x2, \y1 - 12.5) -- (\x2, \y2 -8);

\draw[dashed] let \p1 = (A.two), \p2 = (A.center), \p3 = (C.west) in
  (\x3-7.5,\y2+13) -- (\x1+16.5,\y2+13)
                   -- (\x1+16.5,\y2-17.5)
                   -- (\x3-7.5 ,\y2-17.5)
                   -- cycle
  ;

\end{tikzpicture}
\quad
\begin{tikzpicture}[list/.style={rectangle split, rectangle split parts=2,
  draw, rectangle split horizontal, rectangle split draw splits=false}, >=stealth, start chain,
  node distance=.20cm]

\node[on chain] (C) {\framebox{$\sinj[]2\hphantom{\bullet}$}};
\node[list,on chain] (A) {\,$\mathllap{(}$\hspace{10pt}\nodepart{two}\hspace{-3pt},\hspace{4.5pt}$\mathrlap{)}$\,};
\node[on chain] (B) {\framebox{$42$}};

\draw[*->] let \p1 = (C.center), \p2 = (A) in (\x1 + 7, \y1) -- (A);
\draw[*->] let \p1 = (A.two), \p2 = (A.center), \p3 = (C.west) in
(\x1,\y2) -- (\x1 + 15, \y2) -- (\x1 + 15, \y2 + 10) -- (\x3 - 5, \y2 + 10)
          -- (\x3 - 5, \y2) -- (\x3+4, \y3);
\draw[*->] let \p1 = (A.one), \p2 = (B) in (\x1+3,\y1 + 5) -- (\x1 + 3, \y1 - 12.5) -- (\x2, \y1 - 12.5) -- (\x2, \y2 -8);

\draw[dashed] let \p1 = (A.two), \p2 = (A.center), \p3 = (C.west) in
  (\x2-18,\y2+13) -- (\x1+40,\y2+13)
                   -- (\x1+40,\y2-17.5)
                   -- (\x2-18 ,\y2-17.5)
                   -- cycle
  ;
\end{tikzpicture}
\end{center}

  On the right we depict a representative for a value in
  $P\stores\set{\ell:\linkedlist}$, whose private locations are given
  by $\set{\ell_0 : \cell, \ell_1 : \data}$.
\end{example}

The contravariant action of the coend gives, for every injection
$\h : \w_1 \to \w_2$, a function \( \mhide_\h : PA\w_2 \to PA\w_1 \)
defined by $q_{\h' : \w_2 \to \w'}(a) \mapsto q_{\h'\compose\h}(a)$
(\subfigref{hiding hide}).  The unit is given by
$\mreturn^P_\w \definedby q_{\id[\w]} : A\to PA$ (\subfigref{hiding
  return}).  For every $g : A \to PB$, define $\bind g : PA \to PB$ by
(\subfigref{hiding bind})
\[
(q_{\h : \w \to \w'}(a)\bind g) \definedby \mhide_{\h}(g_{\w'}(a))
\]
For every initialisation $\ih : \w_1 \to \w_2$, we derive the functorial
action $PA\ih : PA\w_1 \to PA\w_2$ (\subfigref{hiding functorial}):
\[
PA\ih(q_{\h : \w_1 \to \w'}(a)) \definedby q_{\ih\mis\h}(A(\h\mis\ih)(a))
\]
Finally, for every $X \in \vals$ and $A \in \pconfs$, define the strength:
\[
\strength(x, q_{\h : \w \to \w'}(a)) \definedby q_{\h}(\pair{X\h x}a)
\]
\begin{proposition}
  The data $\structure P = \seq{P, \mreturn, \bind, \strength}$
  define a strong monad over the $\vals$-actegory $\pconfs$.
\end{proposition}

As a consequence of \propositionref{global state monad transformer},
we obtain a monad over $\vals$, and further calculation using the
explicit description of $\acthom$ shows this monad is the monad
$\structure T$ for full ground storage from the previous section.

\subsection{Hiding algebras}

To shed some light onto $P$, we characterise its algebras. We define a
\emph{hiding algebra} $\structure A = \pair A{\mhide^{\structure A}}$ to consist of a
functor $A \in \pconfs$, and for every morphism $\h : \w_1 \to \w_2$
in $\worlds$, a function $\mhide^{\structure A}_\h : A\w_2 \to A\w_1$,
such that: $\mhide^{\structure A}_{\id[\w]} = \id[A\w]$
for any $\w \in \worlds$;
$\mhide^{\structure A}_{\h_1} \compose \mhide^{\structure A}_{\h_2} =
\mhide^{\structure A}_{\h_2 \compose \h_1}$ for every two composable arrows
$\h_1$, $\h_2$ in $\worlds$; and
whenever we have two initialisations $\ih_1 : \w_1 \to \w_2$,
$\ih_2 : \w_3 \to \w_4$ in $\init$ and two injections
$\h_1 : \w_1 \to \w_3$, $\h_2 : \w_2 \to \w_4$ in $\worlds$ such that
\begin{itemize}
\item $u\ih_2 \compose \h_1 = \h_2 \compose u\ih_1$, i.e., a commuting square;
\item for every $\ell_2 \in \w_2$ and $\ell_3 \in \w_3$, if
  $\h_2(\ell_2) = \p\ih_2(\ell_3)$ then there exists some (necessarily
  unique) $\ell_1 \in \w_1$ such that $\p\ih_1(\ell_1) = \ell_2$ and
  $\h_1(\ell_1) = \ell_3$; and
\item for each $(\ell_2 : \Cll) \in \w_2\o-u\ih_1$, taking the unique
  $\ell_3$ in $\w_4\o-u\ih_2$ such that $\h_2\compose
  \comp{(\p\ih_1)}(\ell_2) = \comp {\p\ih_2}(\ell_3)$, we require that
  $\store_{\ih_2}(\ell_3) = \sem{\typeof
    \Cll}(\h_2)(\store_{\ih_1}(\ell_2))$,
\end{itemize}
then we have the equation
$A\ih_1 \compose \mhide^{\structure A}_{\h_1} = \mhide^{\structure
  A}_{\h_2} \compose A\ih_2$.
A \emph{hiding homomorphism} $\structure A \to \structure B$ is a
natural transformation $\alpha : A \to B$ of functors in $\pconfs$
such that for every $\h : \w \to \w'$ in $\worlds$, we have the
equation $\mhide^{\structure A}_{\h} \compose \alpha_{\w'} =
\alpha_{\w} \compose \mhide^{\structure B}_{\h}$.

The third hiding algebra axiom has the following computational
intuition. The premise of the third axiom states that the newly
allocated locations in $\w_2$ are disjoint from the private locations
in $\w_3$, and that the initialisation data from $\ih_2$ of public
locations in $\w_4$ do not access any private data, and can be
promoted from the initialisation data in $\ih_1$. The third
requirement then states that this additional data can be encapsulation
by the hiding operation. When applied to program configurations
$X \. \stores$, this condition will give our monad garbage collection
capabilities.

Let $\HidingAlgs$ be the category of hiding algebras and their
homomorphisms, and $\pconfs^{\structure P}$ the category of
$P$-algebras and their homomorphisms. We have an evident forgetful
functor $U$ from each of those categories into $\pconfs$.

\begin{theorem}
  Mapping an Eilenberg-Moore algebra $\pair A\alpha$ to the hiding
  algebra $\pair A{\mhide_{\h} : A_{\w_2} \xto{q_{\h}} PA\w_1
    \xto{\alpha_{\w_1}} A\w_1 }$ is the object part of an isomorphism $\Phi :
  \pconfs^{\structure P} \isomorphic \HidingAlgs$ satisfying $U
  \compose \Phi = U$.
\end{theorem}

\subsection{Effect masking}
To evaluate the monad $T$, we show it can mask hidden effects. First,
we define a semantic criterion for not leaking any locations. We say
that a functor $X \in \vals$ is \emph{constant} when, for every $\h :
\w \to \w'$ in $\worlds$, the function $X\h$ is a bijection. We say
that a world $\w$ is \emph{constant} if, for every $(\ell : \Cll) \in
\w$, the functor $\sem{\typeof\Cll}$ is constant. When $\w$ is
constant, so is every sub-world $\hat\w \to \w$, and the covariant
action of the partially applied functor $\prostores(\w, \placeholder)$
is a natural isomorphism. Given $\h : \w \to \w'$, we can then project
any heap in $\stores\w'$ to a heap in $\stores\w$.
\begin{lemma}\lemmalabel{invertible unit}
  If $\w$ is constant, then the monadic unit is invertible
  $\mreturn^P_{\stores} : \stores\w \xto{\isomorphic} P\stores\w$. In particular,
  $P\stores\emptyset \isomorphic \terminal$.
\end{lemma}
We use this result when we prove the Effect Masking \theoremref{effect
  masking}, as well as when working with concrete examples.
\begin{lemma}\lemmalabel{invertible strength}
  For every constant functor $X \in \vals$ and every $A \in \pconfs$,
  the tensorial strength $\strength^P_{X, A}$ is an isomorphism.
\end{lemma}
While technical, this last result is useful, as it plays the role of
the mono
requirement~\cite{moggi:computational-lambda-calculus-and-monads} in
$\lamref$'s adequacy proof.

We can now prove that morphisms that do not leak locations are
denotationally equivalent to pure values:
\begin{theorem}[effect masking]\theoremlabel{effect masking}
  For every pair of constant functors $\Gamma, X \in \vals$, every
  morphism ${f : \Gamma \to TX}$ factors uniquely through
  the monadic unit: \cdiagram{effect-masking-01}
\end{theorem}

The proof of this theorem is conceptually high-level using our
decomposition of $T$ as $\stores \acthom P(\placeholder\.\stores)$:
\begin{proof}[sketch:]%
  As $\Gamma$ is constant, it suffices to prove the theorem for
  $\Gamma = \terminal$.
  \begin{figure}
  \[
  \begin{adjunctions}[symbol=\longrightarrow]
    \terminal & \stores \acthom P(X\.\stores) & \text{ in $\vals$}\\
    \stores & P(X\.\stores) & \text{ in $\pconfs$}\\
    \stores & X\.P\stores   & \text{ in $\pconfs$, by \lemmaref{invertible strength}}\\
    \init(\emptyset, \placeholder) & X\.P\stores &  \text{ in $\pconfs$}\\
    \terminal &
    (X\.P\stores)\emptyset   & \text{ in $\Set$, by Yoneda}\\
    \terminal & X\emptyset & \text{ in $\Set$, by \lemmaref{invertible unit}}\\
    \terminal & X & \text{ in $\vals$}
  \end{adjunctions}
\]
  \caption{High-level proof of the Effect Masking \theoremref{effect masking}}\figlabel{proof}
  \end{figure}
  Calculate as in \figref{proof}, chasing a generic morphism
  upwards.\nobreak
\end{proof}

We named the factored morphism $\runST f$ as we can use it to
interpret a monadic
metalanguage~\cite{moggi:computational-lambda-calculus-and-monads}
containing a construct similar to Haskell's
$\runST$~\cite{launchbury-spj:lazy-functional-state-threads}.

As usual in functor categories, two different functors may have the
same global elements. Thus, even if $TX$ has the same global elements
as $X$, for any constant $X$, the two functors might differ, for
example, for $X = \terminal$ and the signature from
\exampleref{ex1}. The fact that $T\terminal \not\isomorphic \terminal$
for this signature will be an immediate consequence of $\lamref[]$'s
adequacy (see \exampleref{counter example} below).  However,
computations that do not assume anything about the heap nor leak
references are pure:
\begin{proposition}\propositionlabel{constant at 0}
  For every constant $X \in \vals$, we have $\mreturn^T_{X} :
  X\emptyset \xto\isomorphic TX\emptyset $.
\end{proposition}
To see why it holds, note that the initiality of $\emptyset$ in $\worlds$ means
we can bijectively turn an arbitrary element of $TX\emptyset$ into a
global element. We then bijectively apply effect masking to get a
global element of $X$, equivalently an element of $X\emptyset$, and
further calculation shows the monadic unit induces it.


\section{Semantics for full ground storage}\seclabel{adequacy}
We now return to the $\lamref[]$-calculus.

\subsection{Semantics}
\figref{type sem} presents the interpretation of $\lamref$'s types as
functors in $\vals$. It extends the interpretation of full ground
types by interpreting function types using the exponentials in $\vals$
and the full ground storage monad $T$.

\begin{figure}
\centering
\begin{mathpar}
      \shade{\sem{\sref\Cll}\w \definedby \set{\ell \in \locs \suchthat (\ell : \Cll) \in \w }
        \qquad
      \sem{\sref\Cll}\h(\ell) \definedby \ell}

  \sem{\s0} \definedby \initial

  \sem{\ty_1 \s+ \ty_2} \definedby \sem{\ty_1} + \sem{\ty_2}

  \sem{\s1} \definedby \terminal

  \sem{\ty_1\s*\ty_2} \definedby \sem{\ty_1} \times \sem{\ty_2}

  \sem{\ty_1 \sto \ty_2} \definedby (T\sem{\ty_2})^{\sem{\ty_1}}

  \sem{\Tys} \definedby \prod_{(x : \ty) \in \Tys} \sem\ty
    \end{mathpar}
  \caption{Type semantics}\figlabel{type sem}
\end{figure}

\begin{figure}
  \begin{mathpar}
      \shade{\vsem\ell(\h,\me) \definedby\h(\ell)}
\quad
      \vsem\x(\h,\me) \definedby \me(\x)

      \vsem{\sinj i\sv}(\h, \me) \definedby \injection_i(\vsem\sv(\h, \me))
\ \
      \vsem{\sunit}(\h, \me) \definedby \star

      \vsem{\spair{\sv_1}{\sv_2}}(\h, \me) \definedby \pair{\vsem{\sv_1}(\h, \me)}{\vsem{\sv_2}(\h, \me)}

      \vsem{\sfun\x\ty\st}(\h, \me) \definedby \curry\sem{\st}(\h, \me)
    \end{mathpar}
\caption{Value semantics}
  \figlabel{val sem}
\end{figure}

\begin{figure}
  \begin{mathpar}
      \sem{\sinj i\st}(\h, \me) \definedby T\injection_i(\sem\st(\h, \me))

      \sem{\spair{\st}{\st'}}(\h, \me) \definedby
      \dstrength\pair{\sem{\st}(\h, \me)}{\sem{\st'}(\h, \me)}

      \sem{\szmatch \st\ty} = []

      {\sem{\setlength{\jot}{-10pt}
        {\begin{aligned}
            &\sbmatch [{{}\\&\quad}]
                     {\st}
                     {\x_1}{\st'_1\\[0pt]&\quad}
                     {\x_2}{\st'_2}
        \end{aligned}}
      }{\setlength{\jot}{-10pt}
          \begin{aligned}\\
            (\h, \me) \definedby
            \sem\st(\h, \me)\bind\lambda \injection_ia.\\\sem{\st_i}(\h, \me[\x_i \mapsto a])
            \end{aligned}
      }}

      \begin{aligned}
      \sem{\spmatch \st {\x_1}{\x_2}{\st'}}&(\h, \me)
      \definedby\\[-10pt]
      &\mspace{30mu}\strength(\pair\h\me, \sem{\st}(\h, \me))
      \bind
      \sem{\st'}
      \end{aligned}

      \sem{\st\ \st'}(\h, \me) \definedby \dstrength(\sem{\st }(\h, \me),
                                                    \sem{\st'}(\h, \me))
                                         \bind \eval

      \shade{\sem{\sset{\st}{\st'}}(\h, \me)
        \definedby
          \dstrength(\sem{\st }(\h, \me),
                     \sem{\st'}(\h, \me))
          \bind\mset}

       \shade{\sem{\sget{\st}}(\h, \me)
        \definedby
        \sem{\st }(\h, \me)
        \bind\mget}

        \shade{\setlength{\jot}{-10pt}
          \sem{\!\!
            \begin{aligned}
              &\snew {            \\
              &\
                \newbind {\x_1}{\Cll_1}\sv_1\newsep\\
              &\ \vdots                            \\
              &\
              \newbind {\x_n}{\Cll_n}\sv_n          \\
              &}{\st}
            \end{aligned}
          }
          \!\!
          \mathrlap{
          \begin{aligned}[b]
            &(\h, \me)\\
            &\definedby\\\vphantom{\mnew}
          \end{aligned}}
          \begin{aligned}[t]
            &\quad\strength(\pair\h\me,
            \mnew_{\set{\ell_1 : \Cll_1, \ldots, \ell_n : \Cll_n}}\\
            &\qquad{\seq[i=1][n]{\vsem{\sv_i}\!(\h, \me\coseq[i=1][n]{\x_i \mapsto \iinj_2\ell_i})}}\big)
            \\&\quad\bind \sem{\st}
          \end{aligned}
        }
    \end{mathpar}
\caption{Term semantics}\figlabel{term sem}
\end{figure}

We define two semantic functions, for values in \figref{val sem} and
for terms in \figref{term sem}, by induction on typing
judgements. These functions have the following types:
\[
\begin{array}{*2{@{}l}}
\vsem{\Tys\infer_{\w} \sv : \ty} &: \worlds(\w, \placeholder)\times \sem\Tys \to
\hphantom{T}\sem\ty
\\
\sem{\Tys\infer_{\w} \st \,: \ty} &: \worlds(\w, \placeholder)\times \sem\Tys \to
T\sem\ty
\end{array}
\]
The two semantic functions relate by
\(
{\sem{\sv} = \mreturn^T \compose \vsem{\sv}}
\)
and consequently we omitted from \figref{term sem} the definitions
implied by this relationship. The two interpretations take as argument
a \emph{location environment} $\h$, assigning a location in the
current world for every location in the heap layout the term assumes,
and the more standard \emph{(identifier) environment} $\me$, assigning
a value of the appropriate type to every identifier in the type
context.

The definition makes use of the symmetry, dual strength, and the
double strength morphisms:
\[
\begin{array}{*4{@{}l}}
  \swap &{}\definedby \pair{\projection_2}{\projection_1}
  &{}: X\times Y  &{}\to Y\times X
  \\
  \strength' &{}
  \definedby T\swap \compose \, \strength \compose \swap
  &{}: (T X)\times Y &{}\to T(X\times Y)
  \\
  \dstrength &{}
  \definedby (\bind\strength) \compose \strength'
  &{}: (T X)\times(T Y) &{}\to T(X\times Y)
\end{array}
\]
The double strength is given explicitly by
\[
(\projection_{\h_1}\dstrength(\alpha, \beta))(\store_1) = q_{\h_3\compose\h_2}(\pair{X\h_3x}{y}, \store_3)
\]
where $(\projection_{\h_1} \alpha)\store_1 = q_{\h_2}(x, \store_2)$
and $(\projection_{\h_2\compose\h_1}\beta)\store_2 = q_{\h_3}(y,
\store_3)$.

The value semantics is standard, with locations interpreted by the
location environment. The term semantics is standard. The
interpretation of the empty match construct is given by the empty
morphism $[] : \initial \to T\sem\ty$, as having a morphism $\sem\st :
\worlds(\w, \placeholder) \times \sem \Tys \to \initial$ necessitates
$\worlds(\w, \placeholder)\times\sem\Tys$ is isomorphic to
$\initial$. The interpretations of the three storage operations use
the corresponding three operations for the monad $T$ from
\secref{monad}. There are two steps in defining the semantics of
allocation. First, we interpret the initialisation data in the world
extended with $\w_0$, which gives us the appropriate input to the
$\mnew$ morphism from \secref{monad}. The morphism $\mnew$ then
returns the newly allocated locations, which we bind to the remainder
of the computation.

The semantics satisfies the usual substitution lemma. It is also
uniform with respect to the heap layout in the typing judgement. To
phrase it, note that every layout extension ${\w \leq \w'}$ denotes
the world injection given by inclusion.
\begin{lemma}
  For every layout extension $\w \leq \w'$ we have:
\[
  \begin{array}{l@{\,}ll}
    \vsem{\Tys\infer_{\w'}\sv:\ty}_{\w'}(\id[\w'] , \placeholder)
      &=
    \vsem{\Tys\infer_{\w }\sv:\ty}_{\w'}(\w\leq\w', \placeholder)
      \\
    \sem{\Tys\infer_{\w'}\st\,:\ty}_{\w'}(\id[\w'] , \placeholder)
      &=
    \sem{\Tys\infer_{\w }\st\,:\ty}_{\w'}(\w\leq\w', \placeholder)
  \end{array}
  \]
\end{lemma}
This lemma is the semantic counterpart to the monotonicity of the type
system.

\subsection{Soundness and adequacy}
To phrase our denotational soundness result, we first extend the
semantics to heaps. For brevity's sake, we define the following
notation for \emph{closed} program phrases ${\cvsem{\infer_{\w}\sv}
  \definedby \vsem\sv_{\w}(\id[\w], \star)}$, and
$\csem{\infer_{\w}\st}\definedby \sem\st_{\w}(\id[\w], \star)$. Next,
for every typed heap $\heap \in \heaps\w$ define:
\[
\sem\heap \definedby \seq[(\ell : \Cll) \in
  \w]{\cvsem{\infer_{\w}\heap(\ell) : \typeof \Cll}} \in
\stores\w
\]
The semantic heap operations are compatible with the syntactic heap operations, in the
sense that for every syntactic heap $\heap \in \heaps\w_1$, location $(\ell :
\Cll) \in \w_1$, and value $\infer_{\w_1}\sv : \typeof \Cll$ we have:
\( \cvsem{\heap(\ell)} = \sem\heap(\ell) \) and \(
\sem{\heap[\ell\mapsto \sv]} = \sem\heap[\ell \mapsto \cvsem\sv] \).
For allocation, we need to be more careful. Consider any extension $\w
\leq \w_1$, heap $\heap_1 \in \heaps\w_1$, and fresh locations
$\fresh{\w_1}{\ell_1, \ldots, \ell_n}$. Then let $\w' \definedby \w \+
\set{\ell_1 : \Cll_1, \ldots, \ell_n : \Cll_n}$ and $\w_1' \definedby
\w_1 \+ \set{\ell_1 : \Cll_1, \ldots, \ell_n : \Cll_n}$. Then consider
any initialisation data $\seq[i=1][n]{\infer_{\w'}\sv_i : \typeof
  \Cll_i}$, and let $\ih \definedby
\minit{\seq[i=1][n]{\cvsem{\sv_i}}}$ be the corresponding
initialisation. We then have that \( \sem{\heap_1\coseq[i=1][n]{\ell_i
    \mapsto \sv_i}} =
\stores(\vto{\w_1}{(\w\leq\w_1)\mis\ih}{\w_1'})(\sem{\heap_1}) \).

The operational and denotational semantics agree:
\begin{theorem}[soundness]
  \theoremlabel{semantic soundness} The operational and denotational
  semantics agree: for every closed, well-typed term
  $\infer_{\w}\st:\ty$, extensions $\w \leq \w' \leq \w''$, value $\infer_{\w''}\sv : \ty$ and heaps
  $\heap' \in \heaps\w'$ and $\heap'' \in \heaps\w''$, if
  $\pair\st{\heap'} \bigstep \pair\sv{\heap''}$ then \[
  (\projection_{\w\leq\w'}\csem\st)\sem{\heap'} = q_{\w' \leq
    \w''}({\cvsem\sv},{\sem{\heap''}})
  \]
\end{theorem}
The proof is by induction on typing judgements, using the explicit
description of $T$ given in \secref{monad}.

Given two terms $\judge{\st, \st'}{\ty}$, recall the set
$\Plugged\Tys\w\st{\st'}\ty$ of contexts plugged with $\st$ and $\st'$
from \subsecref{observational equivalence}.

\begin{theorem}[compositionality]\theoremlabel{context compositionality}
  For every pair of plugged contexts
  $\Tys'\infer_{\w'}{\st'_1},{\st'_2}:{\ty'} \in \Plugged\Tys\w{\st_1}{\st_2}\ty$,
  if \(
    \sem{\st_1}=\sem{\st_2}
    \) then \(
    \sem{\st'_1}=\sem{\st'_2}
  \).
\end{theorem}
The proof is by induction on contexts, using the fact that the
semantics is given compositionally in terms of sub-terms.

\begin{theorem}[adequacy]\theoremlabel{adequacy}
  For all terms $\judge{\st_1,\st_2}\ty$, if \(
    \sem{\st_1} = \sem{\st_2}
    \) then \(
    \judge{ \st_1 \ctxeq \st_2} \ty
    \).
\end{theorem}
The proof is standard using the Compositionality and Soundness
theorems. In the final step, where the mono requirement is usually
used, use \lemmaref{invertible strength} to project out the shared
return value of the contexts.

\begin{example}\examplelabel{counter example}
  As promised, we show $T\terminal \not\isomorphic
  \terminal$ in the signature from \exampleref{ex1}. Consider the two program phrases:
  \[
  \begin{array}{r@{}*3{@{}l}}
    \infer_{\set{\ell_0, \ell_1 : \data}}
      & \sunit ,
      & {}\slet* {{}&\cx}{\sget \ell_0 } \\
         &&                         &\sset{\ell_0}{\sget{\ell_1}};\\
         &&                         & \sset{\ell_1}{\cx} \qquad : \s1
  \end{array}
  \]
  We can distinguish the two phrases by dereferencing $\ell_0$. Had
  $T\terminal \isomorphic \terminal$, they would have equal
  denotations, and so the result follows from the Adequacy \theoremref{adequacy}.
\end{example}

\subsection{Program equivalences}
There are fourteen program equivalences (ordinary) ground reference
cells are expected to
satisfy~\cite{levy:global-state-considered-helpful,staton:completeness-for-algebraic-theories-of-local-state},
and Staton has shown they are Hilbert-Post complete. While we do not
check their Hilbert-Post completeness here, we validate them for full
ground references. As some equations require locations to be distinct,
we use the heap layout assumption to avoid aliasing:
\[
  \mspace{-100mu}
  \cbveq{GS6}{\smv_1 : \typeof \Cll_1,  \smv_2 : \typeof \Cll_2}
                        {\set{\ell_1 : \Cll_1, \ell_2 : \Cll_2}}
        {
          \sset{\ell_1}{\smv_1};\\
          \sset{\ell_2}{\smv_2}
        }
        {
          \sset{\ell_2}{\smv_2};\\
          \sset{\ell_1}{\smv_1}
        }
        {\s1}
\]




\section{Conclusions and further work}\seclabel{conclusion}
We gave a monad for full ground references. An important ingredient
was to view the collection of heaps as functorial on
initialisations. Using standard developments in
enrichment~\cite{levy:book,
  egger-moegelberg-simpson:the-enriched-effect-calculus,
  staton:completeness-for-algebraic-theories-of-local-state}, we
decomposed it into a monad for hiding transformed with state
capabilities to better account for subtleties in the monad's
definition. We gave evidence that the monad is appropriate for
modelling reference cells: we showed it yields adequate semantics for
the calculus of full ground references, and also validates the
equations expected from a local state monad, as well as the effect
masking property.

Further work abounds. We would like to use the Effect Masking
\theoremref{effect masking} to account for Haskell's $\runST$
construct~\cite{launchbury-spj:lazy-functional-state-threads} by tying
the denotational semantics derived from said theorem with a more
operational account. We would also like to use our semantics to
investigate the combination of polymorphism and reference cells, as
the issues motivating ML's \emph{value
  restriction}~\cite{wright-simple-imperative-polymorphism} surface
with full ground storage.

We would also like to find monads for general storage, and not just
full ground references. As it is possible to tie Landin's
knot~\cite{landin:the-mechanical-evaluation-of-expressions} with
general references and implement full recursion, we expect to need to
solve some recursive equation to obtain the category of worlds.  It
might be possible to do so with a traditional recursive domain
equation~\cite{levy02:possible-world-semantics-for-general-storage-in-call-by-value},
or using step-indexing
methods~\cite{birkedal-et-al:step-indexed-kripke-models-over-recursive-worlds}.

We would like to find an algebraic presentation for our monad in the
style of Plotkin and
Power~\cite{plotkin-power:notions-of-computation-determine-monads},
and investigate its
completeness~\cite{staton:completeness-for-algebraic-theories-of-local-state}. Doing
so would allow us to account for effect-dependent program
transformations~\cite{kammar-plotkin:algebraic-foundations-for-effect-dependent-optimisations}.
We would also like to give a simpler description of the monad's action
at (full) ground types.  Our decomposition of the full ground
references monad differs from existing decompositions for ground
storage~\cite{staton:completeness-for-algebraic-theories-of-local-state,
  mellies:local-state-in-string-diagrams}. Repeating this
decomposition in the ordinary ground case would lead to new insights
into existing and new models.


\section*{Acknowledgements}
Supported by the ERC grant `events causality and
symmetry --- the next-generation semantics', EPSRC grants EP/N007387/1
`quantum
computing as a programming language'
and EP/N023757/1%
 `Recursion,
guarded recursion and computational effects'
, an EPSRC Studentship, and a Royal Society University Research Fellowship. The
authors would like to thank
Bob Atkey,
Simon Castellan,
Pierre Clairambault,
Marcelo Fiore,
Martin Hyland,
Sam Lindley,
James McKinna,
Paul-Andr\'{e} Melli\`{e}s,
Kayvan Memarian,
Dominic Mulligan,
Jean Pichon-Pharabod,
Gordon Plotkin,
Uday Reddy,
Alex Simpson,
Ian Stark,
Kasper Svendsen,
and
Conrad Watt
for fruitful discussions and comments.


\bibliographystyle{IEEEtranS}

\end{document}